\documentclass[manuscript,screen]{acmart}
\AtBeginDocument{%
  }

\setcopyright{none}
\renewcommand\footnotetextcopyrightpermission[1]{}

\copyrightyear{2018}
\acmYear{2018}
\acmDOI{XXXXXXX.XXXXXXX}
\acmConference[Conference acronym 'XX]{Make sure to enter the correct
  conference title from your rights confirmation email}{June 03--05,
  2018}{Woodstock, NY}
\acmISBN{978-1-4503-XXXX-X/2018/06}

\usepackage{bibentry}   
\nobibliography{sample-base.bib}        
\usepackage{booktabs,array,multirow}
\renewcommand\arraystretch{1.2}
\usepackage{graphicx}
\usepackage{tikz}
\usepackage{pgfplots}
\pgfplotsset{compat=1.18}
\usepackage{xcolor} 
\usepackage{booktabs}   
\usepackage{array}      
\usepackage{ragged2e}   
\usepackage{pdfpages}

\newlength{\MiniBarsUnit} 
\newlength{\MiniBarsBarW} 
\newlength{\MiniBarsGapW} 
\newcommand{\MiniBarsColor}{black}                           

\makeatletter
\newcommand{\MiniBars}[1]{%
  \raisebox{-0.2em}{%
    \begingroup\color{\MiniBarsColor}%
    \def\mb@sep{}
    \@for\mb@v:=#1\do{%
      \mb@sep
      \rule{\MiniBarsBarW}{\dimexpr \mb@v\MiniBarsUnit\relax}%
      \def\mb@sep{\hspace{\MiniBarsGapW}}%
    }%
    \endgroup
  }%
}
\makeatother

\begin{document}
\title{Elderly HealthMag: Systematic Building and Calibrating a Tool for Identifying and Evaluating Senior User Digital Health Software}

\author{Yuqing Xiao}
\email{yuqing.xiao@monash.edu}
\orcid{0000-0003-2922-3614}
\affiliation{%
  \institution{Monash University}
  \city{Melbourne}
  \state{Victoria}
  \country{Australia}
}

\author{John Grundy}
\email{john.grundy@monash.edu}
\orcid{0000-0003-4928-7076}
\affiliation{%
  \institution{Monash University}
  \city{Melbourne}
  \state{Victoria}
  \country{Australia}}

\author{Elizabeth Manias}
\email{elizabeth.manias@monash.edu}
\orcid{0000-0002-3747-0087}
\affiliation{%
\institution{Monash University}
\city{Melbourne}
\state{Victoria}
\country{Australia}}

\author{Anuradha Madugalla}
\orcid{0000-0002-3813-8254}
\affiliation{%
  \institution{Deakin University}
  \city{Melbourne}
  \state{Victoria}
  \country{Australia}}

\renewcommand{\shortauthors}{Xiao et al.}

\begin{abstract}
Digital health (DH) software is increasingly deployed to populations where many end users live with one or more health conditions. Yet, DH software development teams frequently operate using implicit, incorrect assumptions about these users, resulting in products that under-serve the specific requirements imposed by their age and health conditions. Consequently, while software may meet clinical objectives on paper, it often fails to be inclusive during actual user interaction. To address this, we propose \textbf{\textit{HealthMag}}, a tool inspired by GenderMag designed to help better elicit, model and evaluate requirements for digital health software. We developed HealthMag through systematic mapping and calibration following the InclusiveMag framework. Furthermore, we integrated this with a calibrated version of an existing AgeMag method to create a dual-lens approach: \textbf{\textit{Elderly HealthMag}}, designed to aid requirements, design and evaluation of mHealth software for senior end users. We demonstrate application and utility of Age HealthMag via cognitive walkthroughs in identifying inclusivity biases in current senior user-oriented digital health applications. 
\end{abstract}

\ccsdesc[500]{Software and its engineering~Requirements analysis}
\ccsdesc[500]{Human-centered computing~User models}
\ccsdesc[300]{Human-centered computing~Walkthrough evaluations}

\keywords{HealthMag, Health aspects, SE for Digital Health, Inclusive Mag, inclusive design}

\received{20 February 2007}
\received[revised]{12 March 2009}
\received[accepted]{5 June 2009}

\maketitle

\section{Introduction}\label{sec1}

Digital health (DH) software is being increasingly deployed. Many of its end users live with one or more health conditions that can shape interaction and requirements -- e.g., users may operate while in pain or fatigued, recover from errors under low self-efficacy, and balance privacy concerns against dependence on caregivers or clinicians. However, because these constraints vary across individuals and situations, DH software development teams often hold implicit, incorrect assumptions (e.g., that “unhealthy users are less capable in technology” or will tolerate longer, more complex flows), and products often under-serve the additional requirements that health conditions impose (e.g., plain language for low health literacy, robust recovery under fatigue). The result is software that meets clinical objectives on paper but fails inclusively at the moment of interaction with end users. To make DH systems more genuinely inclusive, teams need a compact, validated lens that (i) models health-related requirements in ways engineers can act on and (ii) detects developer-assumption bias before it reaches patients and caregivers.

Existing approaches only partially address the need to operationalise health-shaped interaction constraints in DH development~\cite{chute2022user, kolasa2020value, richardson2023generic, hamine2015impact, fricker_requirements_2015}. Work on digital-health equity, digital literacy, and clinical studies showing that caregiver support can improve adoption and outcomes helps explain why usability and outcomes differ across populations; however, these strands leave inclusiveness gaps unexamined e.g., how to support users with low proficiency but not assume all users are like that due to their health conditions. Accessibility checklists are essential, yet often too coarse to capture breakdowns arising from co-morbidities and fluctuating symptoms, while condition-specific guidance is too narrow to cover the combinations common among older adults. These checklists and guidelines rarely provide an actionable method for eliciting requirements or diagnosing health-related inclusiveness breakdowns during design and implementation. Generic personas create empathy but rarely encode validated facet endpoints (e.g., what “low continuity of care” looks like at a decision point). Methods such as GenderMag show the value of calibrated walkthrough lenses, but they are not scoped to address health status. Therefore, DH teams still lack a concise, validated, health-aware lens usable in routine walkthroughs, backlog grooming, and design critiques.

We address this gap with the development of a novel `Health Magnifier' -- or \textbf{`HealthMag'} -- a systematic, research-based method that enables software practitioners and UX professionals to find and fix health-related usability issues in DH software. Inspired by GenderMag \cite{burnett_gendermag_2016}, HealthMag provides a compact set of health facets (attribute types and value ranges) that help developers to better translate health status into software requirements and into evaluation prompts that can uncover inclusivity bugs. By specialising cognitive-walkthrough questions to these facets and instantiating facet-aware personas, the method exposes where developer assumptions may embed bias. For example, the tacit belief that “unhealthy users are less capable” or will accept longer, more onerous flows. It also ensures that designs accommodate the additional requirements that health conditions entail while preserving agency, so highly capable users are not constrained by paternalistic defaults. Finally, because it operates on combinations of facets, HealthMag naturally captures UX effects arising from co-morbidities—revealing risks that emerge when, for example, low continuity of care co-occurs with low self-efficacy and high privacy concern.

Inspired by and leveraging the InclusiveMag framework \cite{mendez_gendermag_2019}, we do not fixate on the taxonomy details of health condition per se. Diagnoses are personal and, in many cases, only health providers and end users can authoritatively define them. Instead, we focus on the software-engineering side of the interface: the interaction consequences of health status that are observable and design-relevant—e.g., health self-efficacy, trust, tech proficiency, received support. We treat these as first-class requirements variables and keep the core set small enough to hold in working memory during evaluation, while broad enough to matter across contexts. 
To calibrate our Magnifier methods, we conducted interviews with domain experts in domain experts in human–computer interaction (HCI), software engineering (SE), and digital health (DH) who are likely to use HealthMag. In these interviews, experts were asked to suggest and rank facets based on their patients/end users' UX in DH software. At an actionable level, we focused on older adults (65+) and people living with health conditions, since these groups are both prevalent among DH users and are often affected by age- and health-related interaction constraints. We represented age-related differences explicitly by pairing HealthMag with Elderly AgeMag (a calibrated variation of AgeMag that is focused on older adults only) together as a dual-lens Magnifier method \textbf{Elderly HealthMag}. It contained 3 personas structured with both elderly AgeMag and HealthMag facets that surface risks arising from health status, age, or their intersection. Based on feedback, we refined the personas for flexible reuse across products and study contexts. 

We evaluate our proposed Elderly HealthMag facets and Personas using interviews with a range of software practitioners and domain experts. We then applied the refined dual-lens Elderly HealthMag personas to cognitive walkthroughs of two widely used DH applications (with inclusive-design features), assessing where flows succeed or fail for different Elderly Health facets. In this work we want to answer the following three questions:
\begin{itemize}
  \item \textbf{RQ1} How can we use a systematic, evidence-based HealthMag model to better capture health-related requirements that matter to end users at interaction time?
  \item \textbf{RQ2} How can this proposed HealthMag framework be calibrated and validated through expert evaluation so that its facets, endpoints, and questions are usable and discriminative?
  \item \textbf{RQ3} What implications emerge for designing and evaluating human-centred SE methods for digital-health software for older adults—especially when HealthMag is paired with AgeMag to support dual-lens personas and walkthroughs?
\end{itemize}

The key contributions of this work include:

\begin{enumerate}
    \item \textbf{Foundations for health-aware HCI/SE.} We carried out a detailed synthesis of over 130 studies on how health conditions affect UX in DH software, organised for direct design and evaluation use, and a set of 16 evidence-grounded candidate facets that repeatedly appear in the literature and matter at interaction time;
    \item \textbf{A novel HealthMag framework and its facets.} We propose a final set of 5 calibrated facets and evidence-based insights that connect facet endpoints to likely breakdowns and concrete design moves;
    \item \textbf{Elderly AgeMag facets for older adult Digital Health users.} We propose a calibrated AgeMag variant inspired by the existing Magnifier method (AgeMag) and tailored to older adult users, combining with our proposed HealthMag framework to produce a novel Age HealthMag framework;
    \item \textbf{Actionable Dual-Mag with 3 flexible personas for Elderly HealthMag.} We developed three flexible personas that are judged realistic and useful by domain experts and developers; and
    \item \textbf{Cognitive walkthrough reveals dual-lens bias.} We carry out cognitive walkthroughs of existing digital Health apps targeting ageing users to show that our Age HealthMag enables improved requirements capture and evaluation across health and age lenses for a customised end-user group. Empirical evidence of developer-assumption bias surfaced during these app evaluations, with corresponding design remedies and requirements updates produced.
\end{enumerate}

\section{Background and Related Work}\label{sec2}
\subsection{Health-related Requirements Modeling}
Health-related requirements are a recurrent concern in software for disability support, chronic-disease self-management, and acute-care workflows~\cite{chute2022user, kolasa2020value}. Prior work underscores the need for personalised, well-characterised requirements for users with specific conditions, so that design choices reflect concrete differences in goals, capabilities, and contexts~\cite{richardson2023generic, hamine2015impact}. Within SE and HCI, several studies focus on turning stakeholder needs into implementable artefacts that aid understanding, communication, and reasoning~\cite{fricker_requirements_2015}. For example, Fricker et al.\ propose RE guidelines for mHealth that elicit and model stakeholder requirements using established RE techniques, and distil recommendations for project practice and future research~\cite{fricker_requirements_2015}. Similarly, Wang et al.\ report a digital-health design framework that maps twelve common design challenges to eight strategies, clarifying how deliverables, activities, stakeholders, and constraints should align across design stages to keep patient requirements central~\cite{wang_designing_2024}.

However, these lines of work seldom address how requirement models explicitly support inclusiveness at the point of interaction—i.e., how health status (e.g., low health literacy, low self-efficacy, limited continuity of care) manifests in specific breakdowns within a flow and how teams can detect them systematically during evaluation. Complementary streams emphasise user engagement and participation to strengthen requirement fidelity. The CeHRes-style roadmap for eHealth development articulates an integrated cycle of development, implementation, and evaluation, promoting the iterative translation of stakeholder values and context into technology, repeatedly tested with end users and ecosystem actors~\cite{kip_cehres_2025}. Co-design accounts likewise show how patients, caregivers, and clinicians contribute experiential knowledge that can be transformed into actionable design inputs and implementation tasks~\cite{bird_generative_2021, sanders2008co}.

Taken together, existing RE and participatory frameworks provide process and structure for capturing health-related requirements, but they typically stop short of giving engineering teams an evaluable and actionable lens to (i) model health-specific interaction risks and (ii) surface developer-assumption bias during analysis~\cite{trischler2019co, slattery2020research, zallio2022online}. Our work addresses this gap by operationalising evidence-based health facets into personas and walkthrough prompts that make inclusiveness issues visible and actionable in routine SE practice.

\subsection{GenderMag and the Inclusive Mag Method for Modelling and Evaluating Diverse User Software Requirements}
The InclusiveMag “Magnifier” (Mag) family offers a unique, requirements-modelling approach grounded in software--user interaction~\cite{mendez_gendermag_2019}. Rather than starting from high-level demographics or static checklists, Mag methods derive evidence-based facets with value ranges, instantiate them in research-grounded personas, and then specialise an analytic procedure (typically a cognitive walkthrough) so that each step of a real task is interrogated through those facets. This pipeline turns diversity considerations into traceable requirements at the point of interaction—linking facet endpoints to concrete questions, observable breakdowns, and targeted design remedies—thereby providing a holistic bridge from users’ capacities and contexts to the software behaviours that must support them.

GenderMag is the original, extensively validated Magnifier framework instance. It models gender-clustered individual differences through five problem-solving facets (e.g., risk attitude, self-efficacy, information processing style), brought to life via research-based personas (Abi, Pat, Tim)~\cite{burnett_gendermag_2016}. Analysts walk scenarios step-by-step, asking specialised questions at each subgoal/action and logging inclusivity bugs when a persona’s facet endpoint (e.g., risk-averse) is not supported by the design. Prior studies show that fixes driven by the revealing facet can close observed gender gaps and improve the software for everyone. 
From GenderMag, researchers abstracted the general recipe into InclusiveMag~\cite{mendez_gendermag_2019}. In an early multi-case study, eight teams each generated and applied an InclusiveMag-based method across eight diversity dimensions (e.g., age, SES, literacy, disability-related traits), providing initial evidence of generality and practicality. The method emphasises facet quality (grounded in strong evidence), parsimony (a small set that fits in working memory), and usability for practitioners.

A new SES-Mag (SocioeconomicMag) illustrates how InclusiveMag has recently been developed and used to build a new domain-specific method~\cite{burnett_toward_2024}. Burnett et al.\ define the SES dimension, derive SES facets and value ranges from the literature, instantiate customizable multi-personas that embed those facets, and specialise an analytic process so the facets/personas can be applied to real products~\cite{chikezie_measuring_2025}. The pipeline mirrors InclusiveMag’s steps and demonstrates how personas make facets actionable in evaluation and redesign.

Finally, intersectional HCI work shows how multiple Mag lenses can be composed analytically. Fallatah et al.\ provide a type-theoretic framework for decomposing and recomposing diversity facets so that teams can analyse intersections (e.g., Ethnicity $\times$ Gender) with provable equivalence to unions of single-dimension analyses (modulo human error) and with empirical support in team studies. This positions dual- or multi-lens analyses as practical, resource-saving complements to empirical studies~\cite{fallatah_intersectional_2025}.
Taken together, these threads demonstrate that the Mag approach is a systematically founded, validated, portable way to operationalise diversity-aware requirements: evidence-derived facets, persona instantiation, and specialised analytic questions that surface where designs underserve particular facet endpoints and how to remediate them.

Beyond Mag development and calibration studies, existing research shows that Mag is flexible and portable for surfacing bias and steering design~\cite{burnett_gendermag_2016}. In software engineering contexts, researchers have used Mag personas and walkthroughs to detect and remediate inclusivity problems in real systems and in developers’ practice~\cite{culas2025newcomers, padala2020gender}. For example, studies report using GenderMag to debug inclusivity issues in information architecture and validating the resulting fixes with users, reducing inclusivity bugs in practice~\cite{guizani2022debug, kanij2022new}. The approach also supports automation: AID operationalises GenderMag cues to flag gender-inclusivity issues on OSS project pages, illustrating how Mag findings can be translated into detectors and checklists for engineering teams~\cite{chatterjee2021aid, chatterjee2022inclusivity, vorvoreanu2019gender, shekhar2018cognitive}. Beyond gender, Mag’s mechanics have been adapted to other lenses; for instance, age-focused evaluations of e-commerce flows use Mag-style personas and walkthrough prompts to expose age-related barriers and guide concrete design changes~\cite{mcintosh2021evaluating}.

\section{Methodology}\label{sec3}
\subsection{Our Approach}

\begin{figure} [h]
    \centering
    \includegraphics[width=0.9\textwidth]{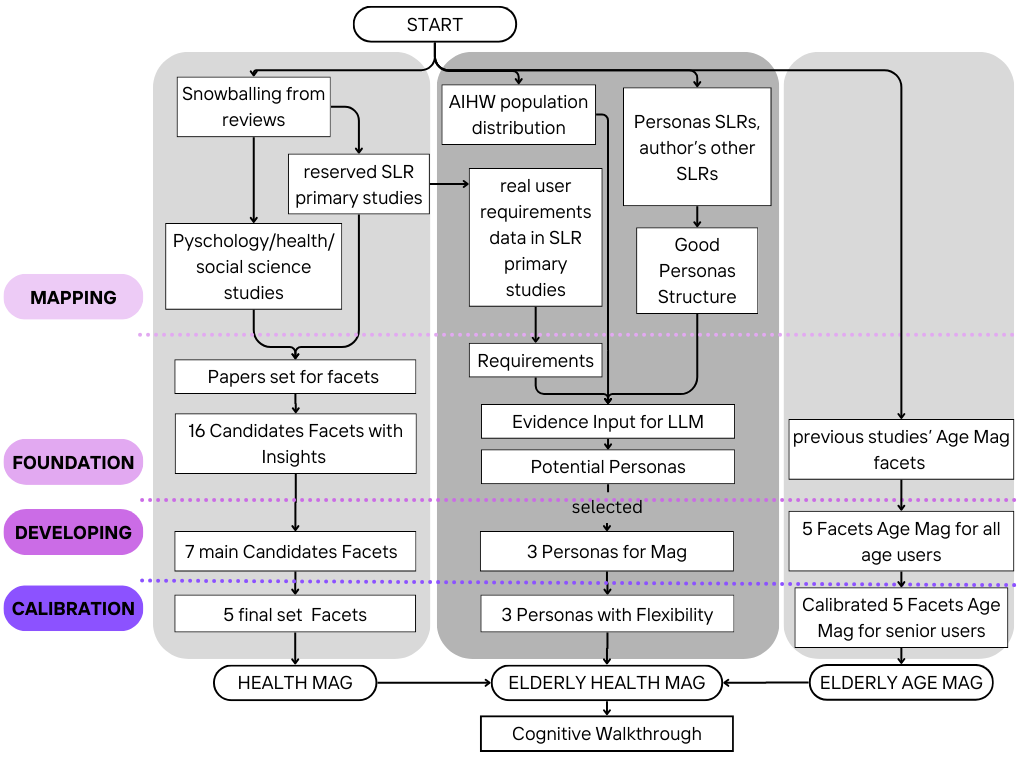}
    \caption{\textcolor{cyan}{The workflow of HealthMag Building, Persona developing \& Calibration} \footnotesize\emph{Note:} Abbreviations—SLR: systemic literature review; AIHW:Australian Institute of Health and Welfare; LLM: Large language model; Mag: (Inclusive) Magnifier.}
    \label{fig:method_process}
\end{figure}

As summarised in Figure~\ref{fig:method_process}, our method for the development of a novel Age HealthMag composite magnifier for ageing user mHealth software development comprises three coordinated research streams. In the left stream, we derive, narrow, and calibrate \textbf{HealthMag} -- a facet set and analysis procedure for users with health conditions. In the right stream, we reuse an existing calibrated \textbf{AgeMag} and calibrate it to a population-specific variant for older adults, yielding \textbf{Elderly AgeMag}~\cite{mcintosh2021evaluating}. For the middle stream, we construct evidence-driven, LLM-generated personas and refine them into flexible, facet-aligned personas for actionable use. The three outputs integrate into \textbf{Elderly HealthMag}, a dual-lens method that pairs HealthMag and Elderly AgeMag with personas to support evaluation and design. Our design scope is to use these Mag in digital health software, but they can be extended to any software for any users with health conditions. 

\textbf{Phase 1: Mapping: } In the left stream we first map the problem space for health by starting from a previous SLR's primary studies and snowballing into digital‑health surveys and psychology, health and social‑science work, yielding a literature set for facet derivation. In the middle stream, we map user requirements by combining AIHW population distributions with real requirements quotes from SLR primary studies and with persona-structure guidance from persona SLRs and our prior SLRs. In the right stream, we map the age dimension by importing facets from previous AgeMag studies.

\textbf{Phase 2: Foundation — assembling evidence} In the left stream, we turn the mapped literature into a set of papers to identify facets. From these we derive 16 candidate health facets with associated insights. In the middle stream, we consolidate population and requirements data into structured requirements and transform them into evidence input for the LLM. In the right stream, we retain the previously identified five AgeMag facets for all‑age users as the starting point~\cite{mcintosh2021evaluating}.

\textbf{Phase 3: Developing — narrowing, insights, and personas} In the left stream, we iteratively narrow the 16 health facets to 7 main candidates and then to a compact final set of 5, constituting HealthMag’s facet core. In the middle stream, the LLM uses the evidence input to generate 20 personas; we then select 3 personas that best span the HealthMag facets for use in the method. In the right stream, the original five AgeMag facets are kept as the working AgeMag facet set.

\textbf{Phase 4: Calibration — expert review and refinement Expert panel ranking} We then conducted calibration sessions with domain experts (clinicians, gerontology specialists, and HCI/SE practitioners). In the left stream, we calibrate the 5 candidate health facets with domain experts, producing our final HealthMag facet set. In the middle stream, we refine our 3 candidate personas into flexible personas that can be tuned while preserving facet integrity. In the right stream, we calibrate the five AgeMag facets specifically for senior users, yielding our `Elderly AgeMag'. We obtained 5 final HealthMag facets with insights, 5 calibrated Elderly AgeMag facets and 3 flexible Health×Age dual-lens personas spanning endpoints and intersections. These three calibrated outputs -- HealthMag, flexible personas, and Elderly AgeMag -- are then combined into our new proposed \textbf{`Elderly HealthMag'}, a dual‑lens method for evaluating and designing software for older adults with health conditions.
This calibration phase was designed to elicit expert feedback on two critical architectural components: the validity of the interaction facets and the representational fidelity of the personas. 

Our calibration protocol proceeded in three logical stages. First, participants were asked to reflect on their professional experience with "age-inclusiveness bugs," establishing the problem context. Second, we presented the candidate facets for HealthMag and Age Mag, asking experts to rank their importance and identify any redundancy or missing factors essential for inspecting digital health apps. Finally, experts reviewed the LLM-generated personas to evaluate their "realism" and identify any stereotypical or "hallucinated medical scenarios" behaviours that might degrade the quality of a cognitive walkthrough.

\subsection{Development of HealthMag}
The development of our novel \textit{HealthMag} was designed to address inclusivity barriers faced by users living with health conditions, including those whose interaction constraints may vary over time and across levels of severity. As illustrated in the left stream of Figure~\ref{fig:method_process}, our process follows a four-phase workflow: \textit{Mapping}, \textit{Foundation}, \textit{Developing}, and \textit{Calibration}.

\subsubsection{Deriving 16 candidate HealthMag facets from the literature (Mapping and Foundation)}
InclusiveMag defines facets as attribute types and value ranges whose values are related to an individual’s diversity dimensions. However, the distribution of these values does not necessarily exhibit a monotonic correlation with a given diversity dimension. That is, combinations of facets can correspond to different diversity dimensions and should not be pre-defined; this increases inclusiveness and allows the model to better meet user needs across diverse situations.
We wanted to find such a set of facets for our focused diversity dimension, the health dimension. Prior research discusses impactful factors that can serve as facets. For example, (i) lower self-efficacy leading to earlier abandonment after ambiguity due to defensive strategies~\cite{zhang2024caregiving, bertolazzi_barriers_2024}, (ii) help-seeking that is relational and caregiver/peer-mediated rather than individual~\cite{czaja2021social, sen2022use}

Accordingly, we defined a Health facet as an attribute type whose values correlate with individuals’ health status in ways that materially affect how they use technology. A facet value is the particular value a person exhibits on that facet (e.g., high vs.\ low health self-efficacy; high vs.\ low continuity with clinicians; high vs.\ low trust in health information). As in InclusiveMag~\cite{mendez_gendermag_2019}, we sought facets whose value ranges are broad enough to meaningfully differentiate experiences and whose endpoints can guide inclusive design: if teams design to support both endpoints, they support the values in between as well.
To achieve this, we derived HealthMag’s candidate facets through a structured, InclusiveMag-style pipeline that links foundations to actionable constructs. Similar to InclusiveMag, we imposed an explicit constraint that facets must be both wide and relevant enough~\cite{burnett_toward_2024}. This ensures they capture a wide spectrum of individuals’ positions and encourages designers to support both ends of each range. Relevance ensures the facets describe technology-use behaviours shaped by health conditions, and width ensures coverage across mild, intermediate, and severe positions.

Alongside this constraint, we used seven review papers as seed studies—spanning clinical, health, social science, and psychology domains—that examine factors shaping the requirements or user experience of people with health conditions, and then conducted snowballing. We ultimately assembled 134 papers (69 in CS/SE/DH and 65 in clinical/health/social science/psychology), including: (1) papers in the software or digital-health domain that investigate how end users’ health conditions are considered in requirements engineering and how they can impact UX for DH software design; and (2) papers in health-related domains (e.g., geriatrics, psychology, behavioural science) showing that patients’ health conditions may affect UX, accessibility, adherence, and health outcomes. Along the way, we incorporated expert recommendations and complemented the corpus with relevant public statistics (e.g., the Australian Institute of Health and Wellbeing) to ground prevalence and variability. Using these studies as a backbone, we clustered concepts, merged near-duplicates, and resolved overlaps to form 16 candidate facets.

\subsubsection{Inclusion criteria for 7 Health Facets (Developing)}
We then applied four inclusion criteria: the factor had to (1) recur across sources, (2) be definable in plain language, (3) add discriminative value beyond existing MAG facets, and (4) be actionable through prompts during walkthroughs. We also ensured balance across facets and kept the set feasible for hand-sized persona cards. This filtering narrowed the pool to seven main candidate facets, preserving coverage while keeping HealthMag lightweight and practical for day-to-day requirements and design activities. We adopted one-word labels for clarity and consistent judgment by non-experts. Each facet spans a \emph{wide range of values} to capture diverse user positions and to encourage designs that support both ends of the range—thereby covering many users in between. Inclusion criteria for a candidate facet:
\begin{enumerate}
    \item C1 (Evidence): At least one scholarly source supports the facet’s potential importance to health‑diverse older adults’ technology use;
    \item C2 (Relevance): The facet clearly implicates design or evaluation choices;
    \item C3 (Range): The facet has a wide range of values across individuals;
    \item C4 (Understandability): The facet is understandable and usable by HCI/SE practitioners without medical specialisation;
    \item C5 (Set‑level): As a set, candidates should be large enough to be impactful but small enough to be cognitively manageable for practitioners (to be narrowed later into a core set).
\end{enumerate}
\subsubsection{Identifying a final five (5) HealthMag facets (Calibration)}
After applying our chosen inclusion–exclusion criteria, the authors reached internal agreement on HealthMag’s seven candidate facets through several rounds of meetings. In these meetings, we drew on our research experience to discuss, rank, and iteratively refine the candidate facets until consensus was reached. Then we conducted an empirical calibration before deploying the method in external evaluations. In contrast to calibrations that focus on defining values of dimension diversity like SESMag did~\cite{chikezie_measuring_2025}, HealthMag’s calibration focuses on operationalising \emph{interaction-time consequences} of health status into a compact, usable, and discriminative facet set. Specifically, calibration targets (i) the content validity of the candidate interaction facets and (ii) set-level coverage and efficiency for time-boxed walkthrough use. Accordingly, we recruited domain experts with substantial expertise and practical experience working with technology users who have health conditions, and who are likely to apply HealthMag in research or design practice.

To calibrate the facets and their definitions, we engaged a multidisciplinary panel of experts ($N=10$). As detailed in Table~\ref{tab:participants_expertise}, the cohort spans the digital health ecosystem, including academic researchers (methodological perspective), gerontology and clinical specialists (domain perspective), and industry practitioners (practical utility and compatibility with requirements-engineering workflows). Since these participants resemble the intended users of HealthMag and the associated personas, their feedback helps ensure that the facet set is both theoretically grounded and usable in practice. The panel also included diverse cultural backgrounds, which supported scrutiny of whether facet definitions and persona cues would generalise across contexts.

We first asked interviewees about their backgrounds and invited them to describe how health conditions affect the user experience of digital health (DH) technologies. We guided them to reflect on which interaction factors were most significant in their context. We then explained our method and presented the seven candidate HealthMag facets with explicit definitions. Experts were tasked with a forced-rank prioritisation of these facets in terms of relevance and expected use, based on their impact on system usability, and were invited to suggest additional facets. Each semi-structured interview comprised a brief pre-interview survey (background and demographics), an interview transcript, and facet-ranking data. We computed each facet’s mean rank (1–7, lower is better) and standard deviation across experts. We retained facets that were consistently ranked in the top five and resolved ties through discussion informed by the interview feedback. Final inclusion decisions were primarily guided by qualitative evidence (e.g., clarity, distinctiveness, and actionability), with mean rank and variability used as supporting signals. Data collection followed a semi-structured interview guideline designed to minimise priming bias while systematically stepping through the method’s components (see Appendix~\ref{app:interview_guide} for the complete instrument).

\begin{table}[!ht]
\centering
\scriptsize
\setlength{\tabcolsep}{1.5pt} 
\renewcommand{\arraystretch}{1.3}
\begin{tabular}{@{} l c c p{0.18\textwidth} p{0.18\textwidth} l c c c c c c @{}}
\toprule
 & & & & & & \multicolumn{6}{c}{\textbf{Expertise Level (1--5)}} \\
\cmidrule(l){7-12}
\textbf{UID} & \textbf{Gen} & \textbf{Age} & \textbf{Current Role} & \textbf{Focus Area} & \textbf{Ethnicity} & \textbf{HC} & \textbf{CP} & \textbf{DH} & \textbf{UX} & \textbf{SE} & \textbf{Acc} \\
\midrule
P1  & F & 25--34 & Research Fellow & Ageing, Digital Health, SE, HCI & East Asian & 4 & 2 & 4 & 4 & 4 & 5 \\
P2  & F & 18--24 & PhD Candidate (\& Pharmacist) & Pharmacy, Clinical Practice & Caucasian & 4 & 4 & 2 & 1 & 1 & 1 \\
P3  & F & 25--34 & PhD Candidate (CS) & UX, SE, HCD & African American & 2 & 1 & 2 & 4 & 4 & 4 \\
P4  & F & 25--34 & Product Dev. Eng. (Data) & Digital Health, Medical Info. & South Asian & 4 & 3 & 4 & 3 & 4 & 2 \\
P5  & F & 45--54 & Chief Consultant (Surgeon) & Geriatric Surgery, Aging & East Asian & 5 & 4 & 3 & 2 & 1 & 1 \\
P6  & F & 45--54 & Chief Consultant (Physician) & Geriatrics, Psych, Nutrition & East Asian & 4 & 4 & 3 & 3 & 1 & 2 \\
P7  & F & 25--34 & Junior Resident (Surgeon) & Clinical Care, Oncology & East Asian & 3 & 2 & 2 & 1 & 1 & 1 \\
P8  & F & 35--44 & Consultant (Physician) & Geriatrics, Aging & East Asian & 4 & 4 & 3 & 3 & 1 & 1 \\
P9  & F & 25--34 & Product Eng. (Requirements) & UX, Human-Centered Design & East Asian & 1 & 2 & 2 & 5 & 3 & 2 \\
P10 & M & 25--34 & PhD Candidate (CS) & Ageing, UX, SE & South Asian & 2 & 2 & 3 & 5 & 5 & 5 \\
\bottomrule
\end{tabular}
\caption{Participant Demographics, Roles, and Self-Rated Expertise. \newline
\textit{Key: HC=Healthcare, CP=Clinical Practice, DH=Digital Health, UX=User Experience, SE=Software Engineering, Acc=Accessibility. Scale: 1 (no knowledge) to 5 (expert).}}
\label{tab:participants_expertise}
\end{table}

\subsection{Development of Elderly Age Mag}
Our \textit{Elderly Age Mag} was intended to help address software inclusivity barriers experienced by older adults. As illustrated in the rightmost column of Figure~1, we follow a systematic four-phase workflow: \textit{Mapping}, \textit{Foundation}, \textit{Developing}, and \textit{Calibration}.

\subsubsection{Age-Inclusive Mag (Mapping and Foundation)}
We used the findings from two existing studies that address age-related inclusivity: InclusiveMag for older adults using email, and AgeMag for e-commerce. In the \textit{InclusiveMag} meta-method~\cite{mendez_gendermag_2019}, an age-inclusiveness method for older adults who use email was proposed, with three facets: \textit{technology comfort}, \textit{attitude toward technology}, and \textit{physical difficulties}. We further draw foundational insights from McIntosh et al.~\cite{mcintosh2021evaluating}, who applied this framework to propose an ``AgeMag'' for eCommerce. Their work identified facets including \textit{technical proficiency and attention}, \textit{risk aversion}, and \textit{vision impairment}, capturing heterogeneous requirements across generations from Generation Z (Gen Z) and Generation Y (Gen Y) to the Silent Generation (SG).

\subsubsection{Refining Existing Mags (Developing)}
Although these foundational studies provide a general lens for age-inclusiveness methods, neither InclusiveMag (email-focused older adults) nor AgeMag (eCommerce across all age ranges) fully matches our target population and context~\cite{mcintosh2021evaluating, mendez_gendermag_2019}. Moreover, preliminary evaluations suggest that participants aged 18--74 (Gen Z to Baby Boomers) exhibited largely homogeneous digital interaction behaviours, whereas the Silent Generation ($75+$) faced distinct barriers---including extreme risk aversion and substantial difficulty with interface legibility---that were not observed in younger cohorts~\cite{mcintosh2021evaluating}. Accordingly, a more targeted model is required to surface these critical nuances. We considered the seven facets reported across both studies, removed overlapping constructs, and, following internal author discussion and agreement, derived six facets: \textit{attention}, \textit{risk aversion}, \textit{vision impairment}, \textit{physical disability}, \textit{trust} and \textit{technical proficiency}.

\subsubsection{Older-Adult-Focused Tool: Elderly Age Mag (Calibration)}
Following the protocol used for HealthMag, we calibrated \textit{Elderly Age Mag} within the same expert interview study by eliciting age-related digital exclusion phenomena observed by domain experts. We initially presented the six candidate facets to the panel to assess their relevance to the heterogeneous capabilities of older adults. Starting from the second interview, experts proposed an additional facet (\textit{education level}), which we added to subsequent interviews for ranking, yielding seven candidates.

We computed each facet’s mean rank (1--7; lower is better) and standard deviation across experts; one participant did not provide a ranking for the added facet and was treated as missing for rank-based statistics. We retained the five facets with the best mean ranks and resolved ties through expert discussion informed by interview feedback. Final decisions were primarily guided by qualitative evidence (e.g., clarity, distinctiveness, and actionability), with mean rank and variability used as supporting signals. Data collection followed a semi-structured interview guide (see Appendix~\ref{app:interview_guide} for the complete instrument).

\subsection{A Novel Composite Elderly HealthMag for Dual-Lens Bias Mitigation}

We wanted to apply our proposed HealthMag with a real target user group of interest to our research group -- older adults with health conditions who use digital health (DH) applications. Many DH users with health conditions are also older adults, and DH applications explicitly designed for older users are often easier to find than those tailored to other dimensions of diversity (e.g., cultural or socio-economic contexts) \cite{khalajzadeh2021modelling,mcintosh2021evaluating,johnson2017designing}. However, this focus also introduces a challenge: age-related interaction constraints can confound the identification of health-related breakdowns during evaluation. To better distinguish health-related issues from age-related barriers—and to make intersectional failures explicit (i.e., breakdowns that arise when health- and age-related constraints co-occur)—we paired HealthMag with an older-adult–calibrated age lens to enable dual-lens analysis. This design strengthens the validity and actionability of the inclusiveness bugs identified during cognitive walkthroughs of DH applications.
We thus wanted to integrate our HealthMag and Elderly AgeMag into a dual-lens method by instantiating facet endpoints in evidence-grounded personas. This supports (i) more concrete facet interpretation during expert interviews (by grounding discussion in plausible user profiles) and (ii) assessment of persona realism and utility for walkthrough-based evaluation.

\subsubsection{Evidence-driven LLM-generated Personas Development (Mapping and Foundation)}
We employed a data-driven generative workflow (Figure~\ref{fig:method_process}, middle column). We synthesised evidence from our Systematic Literature Review (SLR), comprising (i) verbatim quotations from real users and (ii) investigators’ structured requirements distilled into actionable feature lists~\cite{xiao2025requirements}. These were clustered into four health-centric themes (\textit{Medication Support}, \textit{Social Interaction and Emotional Care}, \textit{Emergency Support, Fall Prevention} and \textit{Health and Welfare Monitoring}).

Unlike prior persona-generation approaches, we supplied this large corpus to Large Language Models (LLMs)---specifically ChatGPT-4.0 and DeepSeek~\cite{ouyang2022training, achiam2023gpt, liu2024deepseek}---using a sequence of prompts that strictly constrained the output. The generation was governed by two rigorous inputs: (1) a detailed structure containing the 13 candidate facets (7 from HealthMag and 6 from Elderly AgeMag) we developed for HealthMag and Elderly Age Mag, following the validated digital health taxonomy by Devi et al.~\cite{karolita2023use, karolita2023should}; and (2) demographic distributions from the Australian Institute of Health and Welfare (AIHW) to ensure accurate representation of age, gender, and culturally diverse groups. Using this approach we generated 20 LLM-based persona candidates to obtain a sufficiently diverse pool of plausible profiles grounded in population distributions and literature-derived requirements, since facet-endpoint combinations are combinatorial and cannot be reliably covered by producing only the final personas directly. Generating a larger pool also supports quality control against LLM variability (e.g., stereotyped or internally inconsistent outputs) and enables a transparent, criteria-driven down-selection. We then selected three personas because cognitive walkthroughs are time-intensive; limiting the persona set preserves analysis depth while still spanning key HealthMag and AgeMag endpoints and their intersections. From this set, we selected three personas most relevant to our HealthMag facet endpoints and subjected them to a \textit{Human-in-the-Loop} calibration, as described next.

\subsubsection{Persona Calibration}
To evaluate representational fidelity and operational feasibility, we engaged a multidisciplinary panel comprising SE, HCI, and digital health researchers, alongside industry requirements engineers.  These stakeholders—selected as potential future adopters—reviewed the LLM-generated personas against three rigorous criteria: realism (verisimilitude to real-world populations), functionality (completeness for driving cognitive walkthroughs), and flexibility (adaptability to different software contexts). We incorporated their feedback to refine definitions, adjust facet cues, and remove ambiguities. The result is a set of flexible, actionable personas aligned with the methodological principles of GenderMag and the broader InclusiveMag family. Detailed generation prompts, raw thematic clusters, and the full catalogue of candidate personas are available in Appendix ~\ref{sec:appendix_d}.

\subsection{Cognitive Walkthroughs Using Our Elderly HealthMag}
A cognitive walkthrough (CW) is an expert inspection method for evaluating learnability in which an analyst steps through task scenarios and assesses, at each action, whether a target user would know what to do, notice the correct control, and understand the system feedback~\cite{beer1997pair}. In the InclusiveMag family, CW has been specialised with facet-aware personas to surface inclusiveness bugs during scenario-based analysis~\cite{burnett_gendermag_2016,mendez_gendermag_2019}. We applied our calibrated Elderly HealthMag to conduct structured CWs of two widely used medication-management applications: Apple Health (Medications feature) and Medisafe. The goal was to examine how older adults with different health conditions and ages might experience these systems. We selected these applications for their wide adoption and relevance to core medication tasks (e.g., onboarding, reminders, refills), and because they include explicit inclusive-design features that allow meaningful comparison of where inclusivity succeeds or breaks down.
As summarised in Table~\ref{tab:CW_Tasks}, our walkthroughs covered six essential medication-management tasks that older adults commonly perform: (1) adding a new medication, (2) recording a medication as taken, (3) checking the daily medication schedule, (4) rescheduling a medication time, (5) sharing medication taken records (with caregivers, or family), and (6) changing the app's display language setting.

Before data collection, evaluators reviewed the Elderly HealthMag facets, persona cards, and the CW question set, and completed a short pilot walkthrough to align interpretation of facet cues and logging conventions. Evaluators applied core CW questions (action discovery, control visibility, and feedback interpretability), specialised by Elderly HealthMag’s health and age facets. For each task (and its steps), evaluators recorded (i) task outcome (\textit{Completed}: Yes/No/Partial), (ii) a 1--5 \textit{ease rating} (1 = most difficult, 5 = easiest), (iii) \textit{errors/confusions} (brief descriptions of missteps or unclear UI elements), (iv) estimated time to complete the step/task, and (v) recommendations/comments. For non-positive outcomes (No/Partial) and low ease ratings, evaluators captured on-screen textual or visual evidence and logged candidate inclusiveness bugs, including the associated HealthMag/AgeMag facet endpoint(s) and a concrete design or requirements remedy. After completing the walkthrough for each application, evaluators completed the System Usability Scale (SUS); SUS scores are reported on the standard 0--100 scale.

After completing individual inspections, evaluators convened for a two-hour meeting to compare notes and screenshots to reconstruct the exact screens, visual cues, and decision points each persona would encounter. They then reconciled discrepancies and synthesised an adjudicated bug list with supporting evidence. 

Together, this protocol provides a structured, persona-driven application of Elderly HealthMag to assess how each application supports or challenges older adult users with health-related and age-related interaction constraints.

\begin{table}[ht]
\centering
\footnotesize
\caption{Cognitive Walkthrough Task Definitions for All Personas}
\label{tab:CW_Tasks}
\begin{tabular}{p{0.3cm} p{3cm} p{1.5cm} p{5cm} p{5.5cm}}
\hline
\textbf{Task} & \textbf{Action} & \textbf{Personas} & \textbf{Scenario / User Goal} & \textbf{CW Question} \\
\hline

1 & Add new medication & P1, P2, P3 &
User logs in and opens the medication feature.\newline
User attempts to add a new medication by typing its name. &
Will the user find and open the feature for adding a new medication?\newline
Will the user successfully input the medication name? \\

2 & Record medication taken & P1, P2, P3 &
User attempts to record a medication already scheduled.\newline
User attempts to record a medication that is \textit{not} scheduled. &
Will the user correctly record a scheduled medication?\newline
Will the user record the wrong medication, or add it first before recording? \\

3 & Check today’s medication schedule & P1, P2, P3 &
At 12:00pm, user checks the schedule to see the next due medication. &
Will the user find and open the feature for checking their medication schedule? \\

4 & Reschedule a medication time & P1, P2, P3 &
User attempts to update an incorrect medication reminder time. &
Will the user successfully reschedule the medication reminder? \\

5 & Share medication records with doctor & P1, P2, P3 &
User tries to locate the sharing function and send their medication history. &
Will the user successfully obtain and share their medication record data? \\

6 & Edit Language Settings & P2, P3 &
User wants to change the app language (e.g., to Mandarin).\newline
User checks whether existing medications remain identifiable after switching languages. &
Will the user successfully change the language?\newline
Will they still be able to use the app, or need to switch back and forth due to bilingual preference? \\

\hline
\end{tabular}

\begin{minipage}{0.95\linewidth}
\footnotesize
\textit{Note: Task 6 was not performed for Persona 1 (P1) because P1 is an English-only user and does not require language switching.}
\end{minipage}

\end{table}


\section{Results}\label{sec4}

\begin{figure} [ht]
    \centering
    \includegraphics[width=0.9\textwidth]{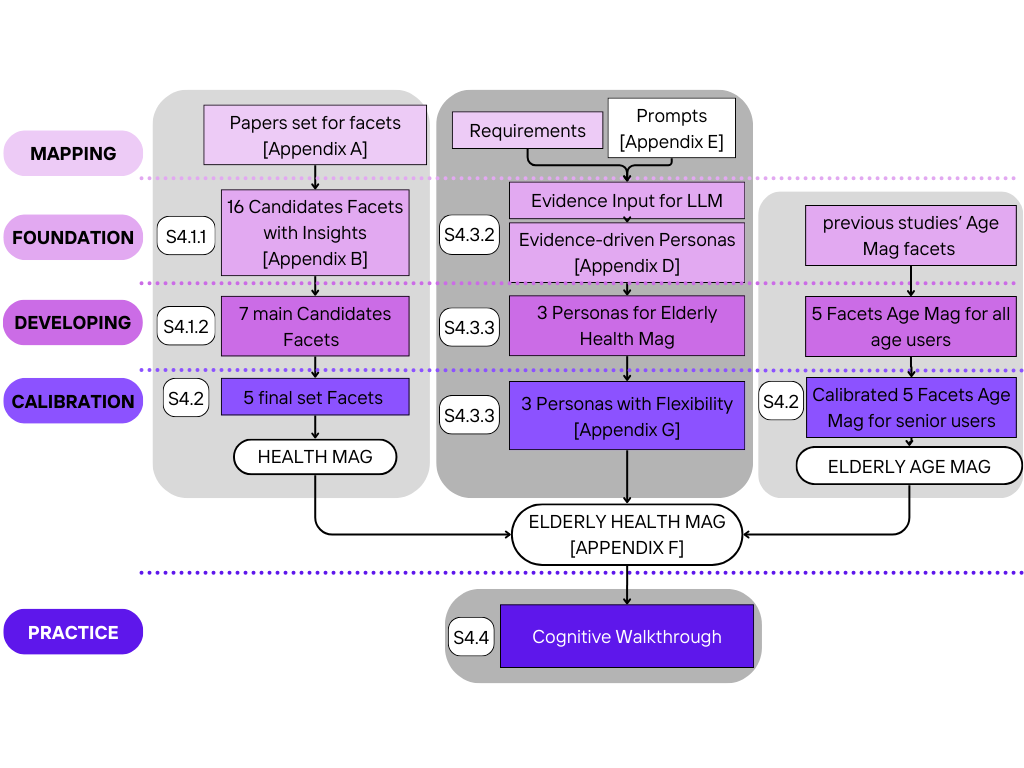}
    \caption{A Summary of Elderly HealthMag Results}
    \label{fig:results_process}
\end{figure}

We describe our production and validation of our new Elderly Health Mag framework following the methods presented in the previous section. As shown in Figure~\ref{fig:results_process} left hand side, to develop HealthMag we first mapped evidence from the literature to derive an initial set of 16 candidate “facets,” then refined these into a seven-facet set and empirically calibrated them into a final five-facet set. On the right side, to develop our Elderly AgeMag, we empirically calibrated the existing Age Mag facets from previous studies to produce a specialised Elderly Age Mag. For our target user population of interest -- older adults who use digital health software -- we translated documented requirements and population distributions into LLM evidence inputs and prompts to generate a broad pool of personas. We then narrowed these down to a set of elderly-specific personas and further adjusted with flexibility considerations. These streams converge into our consolidated Elderly Health Mag framework. Finally, we then evaluated our proposed Elderly HealthMag through a cognitive walkthrough with two widely usaed mHealth apps to assess its utility and alignment with the needs of our intended elderly mHealth software target user audience.

We report our study results in four parts: (i) the final HealthMag facets and associated insights; (ii) the calibrated Elderly AgeMag facet set; (iii) the refined Elderly HealthMag personas; and (iv) findings from applying the dual-lens method in cognitive walkthroughs.

\subsection{HealthMag development}
\begin{figure} [!ht]
    \centering
    \includegraphics[width=0.7\textwidth]{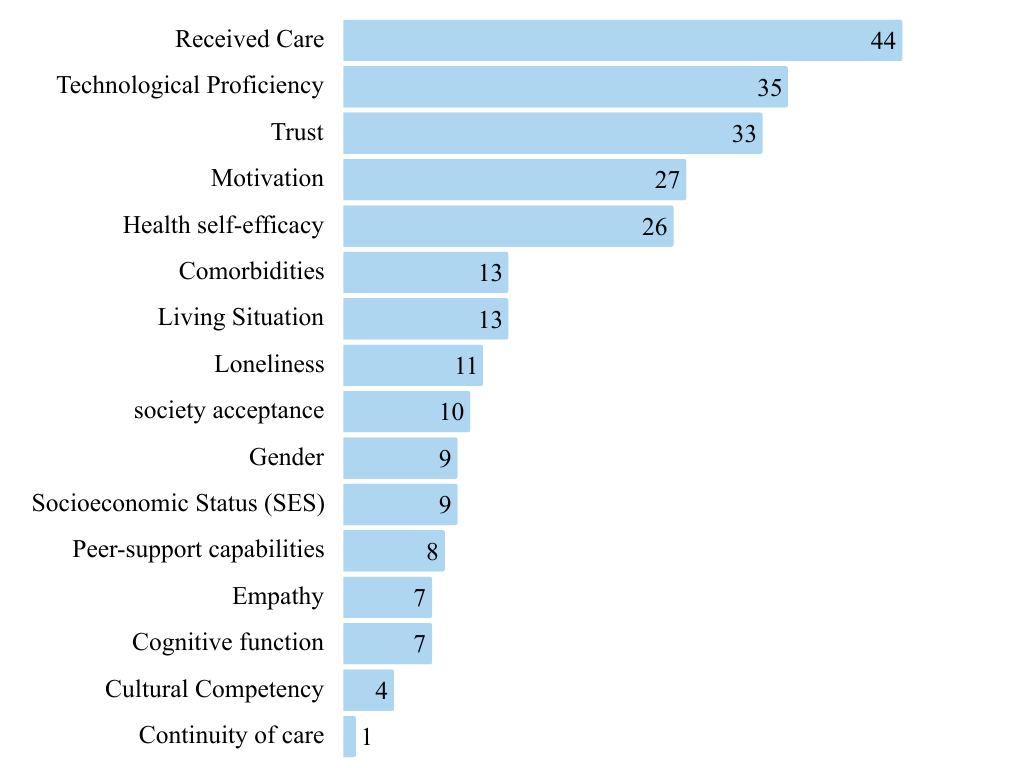}
    \caption{Sixteen candidate facets extracted from the reviewed papers, including the frequency with which each facet is supported in the literature (further details can be found in Appendix \ref{sec:appendix_b})}
    \label{fig:facets_freqs}
\end{figure}

\subsubsection{The 16 Health candidate facets}

From our Mapping, Foundation, and Developing stages (Section 3.1), we derived 16 candidate HealthMag facets from the literature. We treated an attribute as a facet candidate when (i) at least one study provided evidence that it affects technology use for people living with health conditions, (ii) it exhibits a meaningful range of values across individuals, and (iii) it has clear implications for interaction in digital-health software. Figure~\ref{fig:facets_freqs} shows how frequently each candidate appears in our corpus, providing an overview of the evidence base. Across the corpus, the “Received Care” facet appears in 44 papers, while “Cultural Competency” is more emergent—appearing in fewer studies but supported by interaction-relevant evidence; definitions and exemplar impacts are provided in Appendix~\ref{sec:appendix_b}. 

\subsubsection{Seven Candidate Health Facets}
Sixteen candidate facets were evaluated against the inclusion and exclusion criteria outlined in Section 3.2.2, resulting in the selection of seven key facets for HealthMag. The resulting seven facets are: Motivation, Received Care, Tech Proficiency, Cognitive Load, Trust \& Privacy, co-morbidities, and Health Self-efficacy. We elaborate on three that are least captured by existing InclusiveMag dimensions and existing Mag instances. A comprehensive analysis of all calibrated facets and detailed qualitative insights is provided in Appendix~\ref{sec:appendix_c}.

\textbf{Received Care:} The "Received Care" facet captures the extent and quality of assistance a patient obtains from formal sources (e.g., healthcare providers), informal networks (e.g., family), and the broader ecosystem (e.g., government services). We prioritized this facet for two reasons: (1) the level of received care significantly moderates an individual's accessibility to and capability with digital health software; and (2) specific health conditions influence the volume of care a person can elicit from the ecosystem~\cite{albrecht_adherence_2016,baker_primary_2020,barker_association_2017,gray_continuity_2018,haggerty_experienced_2013,hansen_continuity_2013,hemmings_improving_2019, kripalani_reducing_2014,lakerveld_motivation_2020,lorig_self-management_2003,mallick_multivariable_2021}.
This facet spans three dimensions: formal/social care, family support, and peer support. Prior work links social support to higher patient activation~\cite{zhu_relationships_2022} but cautions that worsening conditions can trigger over-assistance that erodes autonomy~\cite{gray_continuity_2018}. Family members often provide setup, reminders, and escalation support, which can be essential for some users~\cite{sorgalla_improving_2017}. Peers can also strengthen coping and confidence, but may introduce misinformation~\cite{gray_continuity_2018}. Our evidence shows wide variance—some older adults have proactive support networks, while others face isolation and fragmented care—so designs that assume “someone else will help” disadvantage users without proxies, and designs that over-rely on proxies can constrain capable users. Therefore, systems should (1) support both self-management and proxy use via configurable roles/permissions and (2) ensure critical tasks are completable independently, with optional, user-controlled ways to invite support.

\textbf{Motivation:} The Motivation facet captures a user’s willingness to initiate and persist with health‑related software over time, whether for monitoring vital signs, post-surgical rehabilitation, or medication management. We selected this facet because (1) an individual's health condition fundamentally impacts their motivation to use digital health tools, and (2) motivation acts as a primary determinant in the decision to adopt specific software.
First, motivation is heavily influenced by the feedback mechanisms inherent in the software, which help maintain user connection~\cite{albrecht_adherence_2016,bertolazzi_barriers_2024,borji_investigating_2018,eisner_influence_2010,haggerty_experienced_2013,kao_association_2019}. For instance, Rodriguez et al. found that in tele-rehabilitation, immediate feedback serves as the key driver of motivation; patients require confirmation that they are executing exercises correctly to sustain engagement~\cite{cleland_contextualizing_2015}. Second, physical health status directly mediates the will to stay healthy. Jackson et al. stated that pain intensity negatively affects the relationship between motivation and function, with high pain levels lowering motivation~\cite{jackson_arthritis_2020}. This aligns with our previous survey study, where older adults reported feeling “too tired,” “discouraged,” or unable to “see the point” due to conditions such as post-surgical pain~\cite{jmirsurveyxiao}. These observations suggest that motivation is not a stable trait but is dynamically shaped by how software frames tasks, reflects progress, and responds to lapses. Consequently, two design imperatives emerge: (1) workflows that visualise progress early and normalise “slips” with easy re‑engagement paths are essential for supporting users with fragile motivation; and (2) features that anchor tasks to personally meaningful outcomes (e.g., maintaining independence) are more effective at sustaining engagement than purely clinical metrics.

\textbf{Health Self‑efficacy:} The Health Self-Efficacy facet reflects an individual's belief in their capacity to manage health tasks, navigate care processes, and execute the behaviours necessary to achieve specific health outcomes~\cite{cleland_contextualizing_2015,albrecht_adherence_2016,hansen_continuity_2013,timmermans_self-management_2024,varsi_implementation_2019,panagioti_self-management_2014,bertolazzi_barriers_2024,bellandi_design_2021,lakerveld_motivation_2020}. In our coding, studies that operationalised health literacy/proficiency primarily as confidence to act under uncertainty clustered with self-efficacy. Meta-analyses indicate that higher health self-efficacy predicts improved self-management and better outcomes across chronic conditions; conversely, low self-efficacy is associated with avoidance, anxiety, and disengagement, even when the individual possesses adequate theoretical knowledge. While related concepts such as "health literacy" and "health proficiency" describe technical capability, we consolidated these under the term "health self-efficacy." This decision stems from the observation that behaviour change in technology usage is driven less by the static capability to process information and more by the user's \textit{belief} in that capability, which varies significantly based on education and life experience. Empirical evidence supports this distinction; for example, self-management interventions have been shown to be more effective than routine care in managing chronic diseases, significantly improving patients’ quality of life and self-efficacy while reducing depressive symptoms~\cite{huang_effect_2024}. Similarly, Jackson et al. report that in cases of OA/RA, higher self-efficacy correlates with better pain and functional outcomes~\cite{jackson_arthritis_2020}. In our systematic literature review (SLR), primary studies revealed that older adults with low efficacy often expressed fear of "getting it wrong" when adjusting dosages or sending messages, interpreting errors or unclear feedback as evidence that they “should not be doing this alone”~\cite{xiao2025requirements,ferreira_elderly_2014,teixeira_design_2017}. These patterns suggest two critical design insights: (1) interfaces should provide scaffolded, low‑risk “practice” environments with clear, non‑blaming error recovery mechanisms; and (2) timely, specific feedback that acknowledges successful steps—rather than focusing solely on failure—can incrementally build self‑efficacy and reduce abandonment.

\subsection{Calibration findings and outcomes}
\subsubsection{Expert qualitative validation: why HealthMag and Elderly AgeMag are needed}
Across interviews, experts consistently affirmed that both \emph{health status} and \emph{age-related constraints} materially shape interaction in digital health (DH) software, supporting the need for a calibrated inspection lens beyond generic usability or accessibility checks. All experts described cases where condition-related factors altered whether users could complete core tasks, and several framed health as a fundamental determinant of how users engage with technology. As P3 put it, health is intrinsic to ``how they show up in the world,'' and therefore affects their capacity to use software.

\begin{table*}[!ht]
\centering
\caption{Expert Calibration Results and Final Facet Selection for HealthMag ($N=10$)}
\label{tab:health_ranking}
\footnotesize
\renewcommand{\arraystretch}{1.3}
\begin{tabular}{p{3cm} c p{8cm} c c}
\toprule
\textbf{Facet Name} & \textbf{Selected} & \textbf{Definition \& Design Prompt} & \textbf{Mean} & \textbf{SD} \\
\midrule

\textbf{Motivation} & $\checkmark$ & 
\textit{Definition:} Willingness to start and persist with health tasks (e.g., logging vitals) over time. \newline
\textit{Prompt:} What helps this user start, and what keeps them going? & 3.3 & 2.1 \\ \addlinespace

\textbf{Tech Proficiency} & $\checkmark$ & 
\textit{Definition:} Experience with and comfort using devices/apps (skill, acceptance, anxiety). \newline
\textit{Prompt:} How gentle is onboarding, and how forgiving is error recovery? & 3.6 & 1.9 \\ \addlinespace

\textbf{Received Care} & $\checkmark$ & 
\textit{Definition:} Availability of social networks (caregivers, clinicians) for use software. \newline
\textit{Prompt:} Does the design enable proxy help and shared views? & 3.7 & 1.9 \\ \addlinespace

\textbf{Cognitive Load} &  & 
\textit{Definition:} Memory, attention, and emotional state required for interaction. \newline
\textit{(Excluded from final independent facet list)} & 4.1 & 1.6 \\ \addlinespace

\textbf{Trust \& Privacy} & $\checkmark$ & 
\textit{Definition:} Propensity to trust the system, including privacy concerns and transparency. \newline
\textit{Prompt:} What must we explain or disclose for informed consent? & 4.2 & 2.6 \\ \addlinespace

\textbf{co-morbidities} &  & 
\textit{Definition:} Dexterity, vision, and co-morbidities affecting interaction. \newline
\textit{(Excluded from final independent facet list)} & 4.3 & 1.7 \\ \addlinespace

\textbf{Health Self-efficacy} & $\checkmark$ & 
\textit{Definition:} Confidence in managing one’s health tasks and navigating care processes. \newline
\textit{Prompt:} Where does the user need guidance or scaffolding? & 4.3 & 2.5 \\

\bottomrule
\multicolumn{5}{l}{\scriptsize \textit{Note: $\mu$ denotes the mean rank (where 1 = highest priority); SD = Standard Deviation. 'Selected' denotes facets retained for the final HealthMag model}}
\end{tabular}
\end{table*}

Experts also emphasised that health-related barriers are not reducible to a single accessibility setting because users’ responses are heterogeneous and often shaped by behavioural adaptation. For instance, P2 described how arthritis can reduce fine motor control, making touch interaction difficult (``can’t click on all the buttons''), and P10 similarly noted that osteoarthritis can cause swollen fingers that change touch accuracy. Others highlighted that even when an impairment is known (e.g., hearing loss), users may not adopt assistive devices (e.g., hearing aids), requiring designs that do not assume external hardware support (P5). Together, these accounts motivate HealthMag’s goal of operationalising \emph{interaction-time consequences} of health status into actionable facets for walkthrough-based evaluation.

In parallel, experts validated the need for an older-adult-calibrated age lens. Eight experts explicitly linked age-related decline to systematic interaction barriers (e.g., legibility, dexterity, attention, and risk posture). Importantly, the panel stressed heterogeneity: older users vary widely in their ability and confidence, and rigid ``senior modes'' can miss the underlying interaction variables that matter. For example, P9 noted that lower technical understanding can manifest as reluctance to trust system recommendations, affecting adoption even when the interface is otherwise usable.

These findings support our dual-lens design. Using HealthMag alone risks misattributing age-driven barriers (e.g., legibility or dexterity limits) to health status, leading to less precise remedies. Pairing HealthMag with Elderly AgeMag makes it possible to distinguish health-related issues from age-related barriers and to surface intersectional failures that emerge when both constraints co-occur.

\begin{table*}[!ht]
\centering
\caption{Expert Calibration Results and Final Facet Selection for Elderly Age Mag ($N=10$)}
\label{tab:age_ranking}
\footnotesize
\renewcommand{\arraystretch}{1.3}
\begin{tabular}{p{3cm} c p{8cm} c c}
\toprule
\textbf{Facet Name} & \textbf{Selected} & \textbf{Definition \& Design Prompt} & \textbf{Mean} & \textbf{SD} \\
\midrule

\textbf{Visual Impairment} & $\checkmark$ & 
\textit{Definition:} Current visual abilities affecting text, icons, contrast, and layout perception. \newline
\textit{Prompt:} Are text size, contrast, and audio options available? & 1.9 & 0.9 \\ \addlinespace

\textbf{Physical Difficulties} & $\checkmark$ & 
\textit{Definition:} Motor/stamina constraints (dexterity, tremor) shaping input pace. \newline
\textit{Prompt:} Can tasks be done with multimodality and adjustable pace? & 2.6 & 1.6 \\ \addlinespace

\textbf{Tech Proficiency} & $\checkmark$ & 
\textit{Definition:} Experience, confidence, and comfort with devices/apps. \newline
\textit{Prompt:} How gentle is onboarding and error recovery? & 2.7 & 1.2 \\ \addlinespace

\textbf{Risk Aversion}  &  & 
\textit{Definition:} Likelihood of taking risk. \newline
\textit{Prompt:} What makes a user willing to try new features? & 3.5 & 1.6 \\ \addlinespace

\textbf{Attention} &  & 
\textit{Definition:} Focus duration and distraction. \newline
\textit{(Excluded from final independent facet list)} & 3.8 & 1.6 \\ \addlinespace

\textbf{Education} & $\checkmark$ & 
\textit{Definition:} Education level. \newline
\textit{Prompt:} Does the design enable proxy actions and shared views? & 1.3 & 0.6 \\ \addlinespace

\textbf{Care received} & $\checkmark$ & 
\textit{Definition:} Help from caregivers/family for setup and troubleshooting. \newline
\textit{Prompt:} Does the design enable proxy actions and shared views? & \multicolumn{2}{c}{\textit{new added}} \\

\bottomrule
\multicolumn{5}{l}{\scriptsize \textit{Note: 'Care received' was inspired by HealthMag and new added.}}
\end{tabular}
\end{table*}

\subsubsection{Five Health Facets (final set)}
Table~\ref{tab:health_ranking} presents the aggregated prioritisation of the seven HealthMag candidate facets. Based on expert rankings and interview feedback, we retained five facets and excluded \textit{Cognitive Load} and \textit{co-morbidities} from the final independent facet set. Experts showed the strongest agreement on \textit{Motivation} and \textit{Tech Proficiency} as foundational determinants of digital-health engagement. Several experts noted that when users lack motivation or confidence with technology, they may disengage early, meaning downstream barriers may never be encountered in practice. Compared to the retained facets, \textit{Cognitive Load} and \textit{co-morbidities} are often direct consequences of specific conditions; rather than treating them as independent facets, we treat them as cross-cutting context variables that are captured through the remaining facets and through standard accessibility considerations.
\textbf{High variance on Health Self-efficacy and Trust.}
\textit{Health Self-efficacy} and \textit{Trust \& Privacy} exhibited the largest standard deviations. Although their mean ranks were lower, they appeared in the top-three priorities for multiple experts ($n=5$), indicating that these facets are critical in particular contexts. This variance also reflected differences in expert perspective: clinicians and HCI researchers (e.g., P1, P3, P5) emphasised Health Self-efficacy as a barrier to adherence, whereas practitioners with a stronger engineering focus tended to view it as more amenable to UI support.
\textbf{Implications for selection.}
Interview comments suggested that Trust can act as a gatekeeper: for some users, it is an immediate blocker (``either in or out straight away,'' P2), while for others it is less salient. Similarly, the variance in Health Self-efficacy indicates that a mean-only selection rule would under-represent ``trust-sensitive'' and ``guidance-dependent'' user profiles. We therefore retained these facets despite their variability to ensure HealthMag remains discriminative across realistic DH contexts.

\begin{table*}[ht]
\centering
\caption{The Consolidated Dual-Mag Facet Set: Integrating Health and Age Constraints}
\label{tab:dual_mag_facets}
\footnotesize
\renewcommand{\arraystretch}{1.5}
\begin{tabular}{p{3cm} p{2.5cm} p{10cm}}
\toprule
\textbf{Facet Name} & \textbf{Origin Source} & \textbf{Description \& Inspection Criteria} \\
\midrule
\multicolumn{3}{l}{\textit{\textbf{Layer 1: Intrinsic Capacity (The "Can Use" Gate)}}} \\
\textbf{1. Visual Impairment} & Age Mag & \textbf{Sensory Capacity.} Inspects if the UI remains usable under degraded vision (e.g., cataracts, diabetic retinopathy). Checks for high contrast and scalability, ensuring the "entry point" is accessible. \\
\textbf{2. Physical Difficulties} & Age Mag & \textbf{Motor Capacity.} Inspects if interaction mechanics accommodate tremors, pain, or reduced dexterity. Ensures targets are large enough and gestures are forgiving of "physical weakness". \\
\hline
\multicolumn{3}{l}{\textit{\textbf{Layer 2: Health Drivers (The "Want to Use" Gate)}}} \\
\textbf{3. Health Motivation} & Health Mag & \textbf{Intent Driver.} Inspects if the system provides sufficient value to overcome inertia. Critical for determining if the user has the "motive" to install the app or if "low engagement" will lead to immediate dropout. \\
\textbf{4. Health Self-Efficacy} & Health Mag & \textbf{Cognitive Resilience.} Inspects if the system demands prior medical knowledge. Checks if users with low health literacy or those who "blindly follow" doctors can interpret data without "complex medical jargon". \\
\textbf{5. Trust} & Health Mag & \textbf{Institutional Confidence.} Inspects for "profit-phobia" and data anxiety. Distinct from risk aversion; specifically targets users who fear the platform is "always making money" or will abuse their information. \\
\hline
\multicolumn{3}{l}{\textit{\textbf{Layer 3: Contextual Enablers (The "How to Use" Support)}}} \\
\textbf{6. Willingness} & Age Mag & \textbf{Operational Confidence.} Inspects for the fear of error (Risk Aversion). Distinct from Trust; targets users afraid of "breaking" the app or those who require "safety nets" (e.g., back buttons) due to past negative experiences of being deceived. \\
\textbf{7. Tech Proficiency} & \textit{Shared (Merged)} & \textbf{Digital Literacy.} Combines \textit{Education} and \textit{Tech Savvy}. Inspects if the user possesses the mental models to navigate hierarchies. Validates if the design supports users who "don't understand... Siri" or lack "operational expectations". \\
\textbf{8. Received Care} & \textit{Shared (Merged)} & \textbf{Extrinsic Support.} Combines \textit{Social Support} and \textit{Available Support}. Inspects if the system allows "proxy interaction" (caregivers) to bridge gaps in SES, connectivity, or ability. Serves as the fallback for all other deficits. \\
\bottomrule
\end{tabular}
\end{table*}

\subsubsection{Elderly Age Facets (final set)}
Table~\ref{tab:age_ranking} summarises experts’ prioritisation of the Elderly AgeMag candidate facets. The panel established a clear hierarchy, ranking \textit{Visual Impairment} and \textit{Physical Difficulties} highest. Qualitative feedback suggested a dependency: P7 noted that ``if one’s physical condition really isn’t up to par, or if their eyesight is poor, then it’s of no use'', making the application effectively unusable regardless of cognitive intent. \textit{Visual Impairment} achieved the strongest consensus, with P10 describing it as a primary entry point for practitioners addressing accessibility needs, and P6 emphasising that vision decline is more widespread than hearing loss in clinical populations.

Based on interview feedback, we excluded \textit{Attention} and refined \textit{Risk Aversion} into \textit{Willingness}. Experts argued that attention limitations are often mitigable (e.g., P7 noted that users can ``rest for a while'' and resume), whereas risk-related reluctance can be a harder barrier shaped by prior negative experiences (e.g., having ``been deceived'' or ``suffered a loss''), motivating safety nets such as reversible actions and clear recovery paths (e.g., P10’s suggestion of ``back buttons''). Finally, we merged \textit{Education} into \textit{Tech Proficiency}, as experts (e.g., P5) observed that education level primarily manifests as operational confidence and expectations.

These calibrated Elderly AgeMag facets were then combined with the five HealthMag facets to form our Dual-Mag method for older adults using digital-health software. After eliminating structural duplicates (\textit{Tech Proficiency} and \textit{Received Care}), the resulting Dual-Mag comprises eight unique facets representing a minimum viable set for modelling intersections between ageing and health management; the final list is reported in Appendix~\ref{sec:appendix_f}.

\subsection{Elderly HealthMag and dual-lens integration}

\subsubsection{Dual-lens composition result}

Table~\ref{tab:dual_mag_facets} presents our consolidated Elderly HealthMag taxonomy, integrating the calibrated HealthMag and Elderly AgeMag into eight non-overlapping facets: \textit{Visual Impairment}, \textit{Physical Difficulties}, \textit{Health Motivation}, \textit{Medical Self-Efficacy}, \textit{Trust}, \textit{Willingness}, \textit{Tech Proficiency}, and \textit{Received Care}. 
To make the dual-lens inspection actionable, we organised the facets into three layers that reflect interaction-time dependencies. \textit{Intrinsic Capacity} forms the ``can use'' gate (vision and motor capacity). \textit{Health Drivers} form the ``want to use'' gate (motivation, self-efficacy, and trust). \textit{Contextual Enablers} provide ``how to use'' support (willingness to act, digital literacy, and available support). 

During consolidation, we merged structurally overlapping constructs across the two lenses. In particular, we retained shared facets for \textit{Tech Proficiency} (absorbing \textit{Education}/digital literacy cues) and \textit{Received Care} (capturing proxy and ecosystem support), while preserving distinct constructs where experts differentiated them (e.g., \textit{Trust} vs.\ \textit{Willingness}). Appendix~\ref{sec:appendix_f} links these facets to expert-reported barriers across intrinsic/clinical, interaction-level, and extrinsic/contextual layers, illustrating how observed breakdowns trace to the facet-driven inspection criteria.

\subsubsection{Data-driven LLM-generated personas}
As shown in the middle stream of Figure~\ref{fig:method_process}, we generated data-driven personas using LLMs to instantiate the endpoints of the HealthMag and Elderly AgeMag facets for dual-lens walkthroughs. The LLM input comprised two components: (1) evidence data and (2) a specified persona structure.

\paragraph{Evidence data}
Our evidence data has two parts: (1) semi-structured, text-formatted requirements and (2) demographic distributions. We extracted empirical evidence from our SLR primary studies in the form of verbatim, real-user requirement statements reported in prior work (37 papers, totaling $>$20{,}000 words). Because these quotations were reported in peer-reviewed publications, they provide attributable grounding for persona needs and pain points. Through thematic clustering, we condensed this corpus into semi-structured requirements used to generate candidate personas.
We obtained distributions of age, language background, and gender from publicly available datasets published by government departments (AIHW). Since we generated 20 personas, we enforced target distributions in the prompts. Based on AIHW language distributions, we specified: 10 English; 2 Italian; 2 Greek; 2 Mandarin; 1 German; 1 Vietnamese; 1 Cantonese; and 1 Punjabi. Based on AIHW age and gender distributions, we specified: 3 men aged 65--70; 3 men aged 71--75; 2 men aged 76--80; 1 man aged 81--85; 3 women aged 65--70; 3 women aged 71--75; 2 women aged 76--80; 2 women aged 81--85; and 1 woman aged 86+.

\paragraph{Persona structure}
Following the persona taxonomy and domain-based customisation proposed by Devi, we used internal and external layers to describe each persona. The persona structure we decided includes: (i) demographic information (name, age, sex); (ii) current situation (e.g., retired, travel habits, community activities); (iii) personal background (e.g., previous job, living situation, family relationships); (iv) health and wellness (conditions, interventions taken); (v) information-seeking behaviour (skill level and tech-savviness); (vi) attitude toward technology (e.g., smartphone use, interests and concerns about e-health apps); (vii) pain points and motivations; and (viii) a usage scenario describing interaction with the software.

\paragraph{Personas result}
Using the above evidence input and persona template (with prompts documented in Appendix~E), the LLM produced 20 candidate personas (see Appendix~\ref{sec:appendix_d}). 

We then selected three personas for Elderly HealthMag use. Our selection was guided by two considerations. First, because we apply a dual-Mag lens (HealthMag + Elderly AgeMag) in the cognitive walkthroughs, we prioritised personas that are representative and plausible for aged-care digital health software contexts. Second, consistent with the exemplar persona sets used in GenderMag, SES-Mag, and other InclusiveMag-derived methods, we selected three personas to span the intended coverage of Elderly HealthMag: one representing the youngest and healthiest profile (Persona 1 in Appendix ~\ref{sec:appendix_d}), one representing an intermediate profile (Persona 8 in Appendix ~\ref{sec:appendix_d}), and one representing the oldest and least healthy profile(Persona 18 in Appendix ~\ref{sec:appendix_d}).
Before proceeding to the next step, we added details about the medications these personas have been taking and carefully checked the personas to ensure alignment with our proposed facets (pre-calibration version). We also added potential photos for all personas. The details of this version of the three potential personas for Elderly HealthMag can be found in Appendix~\ref{sec:appendix_g}.
\subsubsection{Selected and calibrated personas for Elderly HealthMag}
In calibrating Elderly HealthMag, we used personas for three purposes: (i) to validate the \textit{realism} of LLM-generated personas, (ii) to use personas as a concrete vehicle for presenting facet endpoints and eliciting feedback, and (iii) to test and refine the intended \textit{coverage boundaries} of Elderly HealthMag (i.e., which older-adult health/age situations the method should represent).

During calibration, interviewees generally affirmed the realism of the proposed personas, though their assessments reflected distinct professional lenses. \textbf{Health providers} (P2, P5, P6, P7, P8) validated the personas against clinical experience. For example, P2 noted that the character details ``match the personality you described'' and that ``he looks a lot like a real person... people I interact with are very similar to him.'' P6 similarly stated that ``the situation is real ... and the design aspects are quite realistic.'' P8 confirmed that scenarios such as seniors using smartphones for medical research while remaining sceptical are ``quite realistic.'' \textbf{Digital-health researchers} (P1, P4) corroborated these assessments; P4 identified the second persona as the ``most common'' type in Australia and noted that even the severe health scenarios appeared ``real.''

Personas also worked well as a carrier for presenting Elderly HealthMag facets and eliciting expert feedback. \textbf{RE and product experts} (P3, P9, P10) evaluated the personas primarily in terms of internal logic and practical utility. P3 and P10 described the personas as effective and structurally sound, with P3 noting, ``I think this is good... I totally get this. This is a good persona,'' and P10 emphasising that ``personas are a good empathy tool... it makes it more visual... it makes sense.'' Importantly, when used to operationalise Elderly HealthMag facets, the personas prompted targeted, facet-specific feedback. In particular, P3 and P10 highlighted that the personas enabled them to engage with dimensions such as emotional state and health motivation, which they viewed as adding depth beyond conventional requirements representations.

With respect to coverage boundaries, interviewees provided nuanced feedback on whether the personas adequately span the intended user space. Several participants commented on \textit{prevalence} rather than \textit{plausibility}. P2 and P5 observed that the highly healthy persona (Margaret) represents a ``rare'' subset of older users, while still acknowledging that the persona itself was ``quite reasonable'' (especially after setting aside nationality-specific assumptions). In contrast, the ``middle'' persona (Zhao) was consistently recognised as representative of common older-adult users in both Australia and China (e.g., P4, P7). More critical feedback emerged at the lower end of the spectrum: P9 questioned whether the third persona, given very limited technological literacy, would realistically use a mobile application independently, suggesting that interaction would more often be mediated by caregivers or third parties. Overall, the personas were perceived as spanning a meaningful range, but the lower boundary required refinement to better reflect proxy-mediated use in real-world care settings.

Based on this feedback, we refined the personas’ coverage boundaries and interaction assumptions. Table~\ref{tab:persona_update_feedback} summarises the interview-driven calibration decisions and resulting updates to the Elderly HealthMag personas; full details are provided in Appendix~\ref{sec:appendix_g}.

\begin{figure} [h!]
    \centering
    \includegraphics[width=0.8\textwidth]{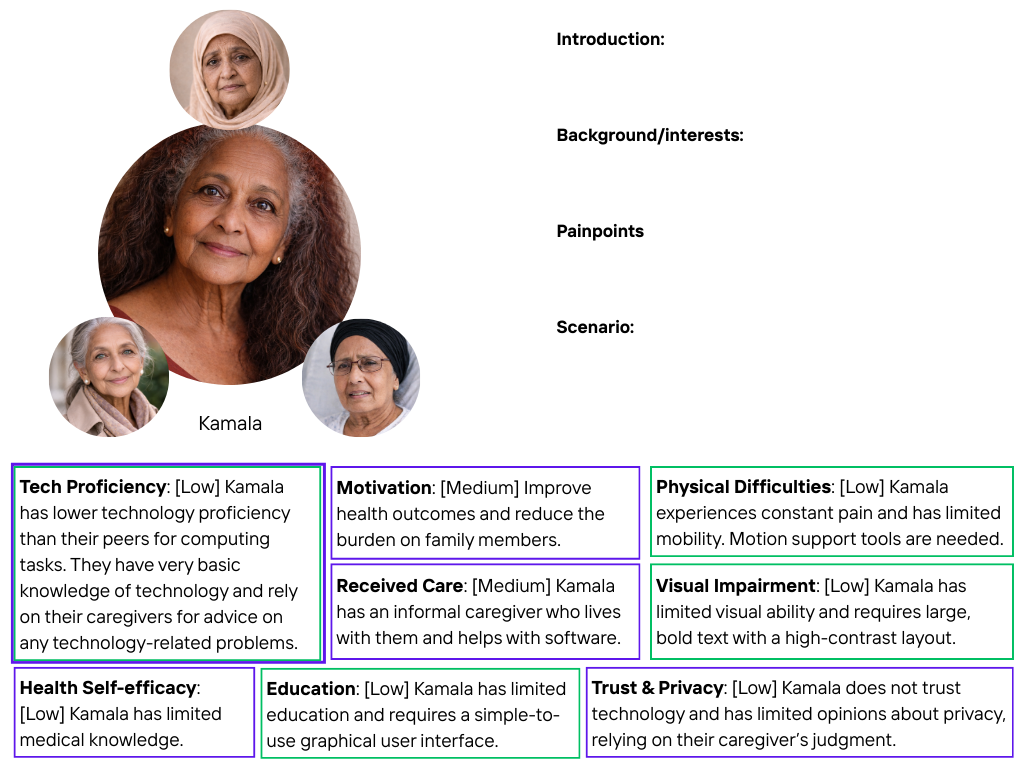}
    \caption{Persona 3: Kamala}
    \label{fig:persona3}
\end{figure}

Figure~\ref{persona3} is one of our 3 personas of Elderly HealthMag. The green square stands for Elderly AgeMag facets, and the purple square stands for HealthMag facets. We left the personal background and other parts of the persona as empty to give flexibility.

\begin{table}[t]
\centering
\caption{Interview-driven persona update suggestions for calibrating Elderly HealthMag}
\label{tab:persona_update_feedback}
\renewcommand{\arraystretch}{1.15}
\resizebox{0.95\textwidth}{!}{%
\begin{tabular}{p{2.3cm} p{7cm} p{4cm} p{2cm}p{2cm}}
\toprule
\textbf{Facet or Persona Component} &
\textbf{Update Suggestion (with Interview Rationale)} &
\textbf{Supporting Quote(s)} &
\textbf{Interviewee \newline (Focus)}& 
\textbf{Related Persona / \newline Addressed } \\
\midrule

Received Care / \newline Tech Proficiency &
Add a \textbf{caregiver- or third-party--mediated persona variant}, reflecting that elderly users with very low technological literacy are unlikely to independently use mobile apps and often rely on family members or caregivers. &
``It might be that \ldots the patient does not use digital health apps, but instead \ldots a third party. For example, here, the son uses it for her.'' &
P4 (DH/RE), \newline P5 (Health) & All personas/ \newline $\checkmark$\\

\midrule
Motivation &
Make \textbf{emotional state and well-being} explicit in persona descriptions, as emotional concerns (e.g., anxiety about ageing and health) were viewed as central to realism and decision-making but underrepresented. &
``I really like emotional state and well-being \ldots it talks about worry that she’s ageing and worrying about her health.'' &
P3 (RE) & 
All personas/ \newline $\checkmark$ \\

\midrule
Socio-economic status &
Add \textbf{socio-economic context} cues (e.g., affordability and device access), acknowledging that economic conditions shape access to technology, priorities, and willingness to engage with digital health tools. &
``Its ease of use \ldots as well as their knowledge level, economic capability, and so on, are all factors involved.'' &
P6(Health), \newline P7(Health), \newline P9 (RE) & 
All personas/ \newline $\checkmark$ \\
\midrule
Coverage of \newline health conditions &
Extend the lower health-severity boundary to \textbf{include more severe conditions}, as the ``least healthy'' persona was considered insufficient to capture situations where self-management is impossible. &
``If a person has very serious complications, they definitely won’t be able to manage it themselves \ldots they rely on others.'' &
P5 (Health) & 
Persona~3/ \newline $\checkmark$ \\

\midrule
Facets structure &
Reduce and refine to \textbf{5-6 facets}, addressing concerns that the original number of facets was too large and overlapping for practical use. &
``I definitely think that’s too many facets \ldots maybe five or six facets would be good.'' &
P3 (HCI/RE) & 
All personas/ \newline $\checkmark$ \\

\midrule
Persona \newline construction &
\textbf{Balance being detailed vs flexibility} for designers, ensuring personas support reasoning without becoming either empty templates or overly rigid representations. &
``It’s important to find a balance \ldots provide some flexibility, but also let them know there are some potential options you can use.'' &
P1 (DH/RE) & 
All personas/ \newline $\checkmark$ \\

\bottomrule
\end{tabular}}
\end{table}

\subsection{Cognitive Walkthrough using Elderly HealthMag}

\begin{table}[!ht]
\centering
\footnotesize
\caption{Developer Roles, Applications, and Assigned Persona--Task Combinations for Cognitive Walkthrough}
\label{tab:CW_evaluators}
\begin{tabular}{p{1.5cm} p{5.5cm} p{2.3cm} p{6cm}}
\hline
\textbf{Developer} & \textbf{Role / Background} & \textbf{Application} & \textbf{Assigned Personas and Tasks} \\
\hline

\textbf{Evaluator 1} & 
Software developer of medication management apps &
Medisafe \newline Apple Health &
\begin{itemize}
    \item Persona 1: Tasks 3, 5
    \item Persona 2: Tasks 1, 2, 3, 5
    \item Persona 3: Tasks 1, 2, 3, 5
\end{itemize} \\

\textbf{Evaluator 2} & 
Software developer of medication management apps &
Medisafe \newline Apple Health &
\begin{itemize}
    \item Persona 1: Tasks 1, 2, 4, 5
    \item Persona 2: Tasks 4, 5
    \item Persona 3: Tasks 1, 2, 4, 5
\end{itemize} \\

\textbf{Evaluator 3} & 
Software developer of medication management apps &
Medisafe \newline Apple Health &
\begin{itemize}
    \item Persona 1: Tasks 1, 2, 3, 4
    \item Persona 2: Tasks 3, 4
    \item Persona 3: Tasks 1, 2, 3, 4
\end{itemize} \\

\textbf{Evaluator 4 \newline(Author 1)} & 
PhD Candidate in Software Engineering; developer of digital health apps &
Medisafe \newline Apple Health &
\begin{itemize}
    \item Persona 1: Tasks 1, 2
    \item Persona 2: Tasks 1, 2, 6
    \item Persona 3: Tasks 1, 2, 6
\end{itemize} \\

\hline
\end{tabular}
\end{table}

\begin{table}[!ht]
\centering
\caption{Cognitive Walkthrough (CW) Tasks with Three Personas, for Two Apps}
\label{tab:cw_tasks_eval_all}
\footnotesize
\renewcommand{\arraystretch}{1.1}
\resizebox{1\linewidth}{!}{%
\begin{tabular}{@{} p{0.10\linewidth} p{0.10\linewidth} p{0.42\linewidth}
                p{0.05\linewidth} p{0.05\linewidth}
                p{0.05\linewidth} p{0.05\linewidth} @{}}
\toprule
 &  &  & \multicolumn{2}{c}{\textbf{Medisafe}} & \multicolumn{2}{c}{\textbf{Apple Health}} \\
\cmidrule(lr){4-5} \cmidrule(lr){6-7}
\textbf{Persona} & \textbf{Task} & \textbf{Description} &
\textbf{Completed} &
\multicolumn{1}{c}{\begin{tabular}{@{}c@{}}\textbf{Ease of Use}\\ \footnotesize(1=hardest, 5=easiest)\end{tabular}} &
\textbf{Completed} &
\multicolumn{1}{c}{\begin{tabular}{@{}c@{}}\textbf{Ease of Use}\\ \footnotesize(1=hardest, 5=easiest)\end{tabular}} \\
\midrule

\multirow{6}{*}{Margaret}
& Task 1 & Add a new medication. &
\multicolumn{1}{c}{Yes} & \multicolumn{1}{c}{4.17} &
\multicolumn{1}{c}{Yes} & \multicolumn{1}{c}{3.83} \\
& Task 2 & Record a medication taken. &
\multicolumn{1}{c}{Yes} & \multicolumn{1}{c}{4.67} &
\multicolumn{1}{c}{Yes} & \multicolumn{1}{c}{4.33} \\
& Task 3 & Check today’s medication schedule (daily reminder). &
\multicolumn{1}{c}{Yes} & \multicolumn{1}{c}{5.00} &
\multicolumn{1}{c}{Partial} & \multicolumn{1}{c}{3.50} \\
& Task 4 & Reschedule a medication time in the plan. &
\multicolumn{1}{c}{Partial} & \multicolumn{1}{c}{3.00} &
\multicolumn{1}{c}{Partial} & \multicolumn{1}{c}{3.50} \\
& Task 5 & Share the medication-taken record. &
\multicolumn{1}{c}{No} & \multicolumn{1}{c}{1.00} &
\multicolumn{1}{c}{Yes} & \multicolumn{1}{c}{4.00} \\
& \multicolumn{2}{l}{\textbf{System Usability Scale, 0--100}} &
\multicolumn{1}{c}{} &
\multicolumn{1}{c}{\textbf{Medisafe SUS:} 76\ \MiniBars{30,30,33,20,23,33,27,33,33,40}} &
\multicolumn{1}{c}{} &
\multicolumn{1}{c}{\textbf{Apple Health SUS:} 73\ \MiniBars{30, 7, 27, 23, 33, 10, 27, 10, 33, 10}} \\
\addlinespace
\midrule

\multirow{7}{*}{Zhao}
& Task 1 & Add a new medication. &
\multicolumn{1}{c}{Partial} & \multicolumn{1}{c}{3.17} &
\multicolumn{1}{c}{Partial} & \multicolumn{1}{c}{2.67} \\
& Task 2 & Record a medication taken. &
\multicolumn{1}{c}{Yes} & \multicolumn{1}{c}{4.00} &
\multicolumn{1}{c}{Yes} & \multicolumn{1}{c}{4.17} \\
& Task 3 & Check today’s medication schedule (daily reminder). &
\multicolumn{1}{c}{Yes} & \multicolumn{1}{c}{5.00} &
\multicolumn{1}{c}{Yes} & \multicolumn{1}{c}{3.50} \\
& Task 4 & Reschedule a medication time in the plan. &
\multicolumn{1}{c}{Partial} & \multicolumn{1}{c}{2.00} &
\multicolumn{1}{c}{Partial} & \multicolumn{1}{c}{2.50} \\
& Task 5 & Share the medication-taken record. &
\multicolumn{1}{c}{No} & \multicolumn{1}{c}{1.00} &
\multicolumn{1}{c}{Partial} & \multicolumn{1}{c}{2.50} \\
& Task 6 & In Settings, change app language to preferred language (Mandarin/Chinese). &
\multicolumn{1}{c}{Partial} & \multicolumn{1}{c}{2.50} &
\multicolumn{1}{c}{Yes} & \multicolumn{1}{c}{4.50} \\
& \multicolumn{2}{l}{\textbf{System Usability Scale, 0--100}} &
\multicolumn{1}{c}{} &
\multicolumn{1}{c}{\textbf{Medisafe SUS:} 61\ \MiniBars{27,27,23,23,23,27,20,30,23,20}} &
\multicolumn{1}{c}{} &
\multicolumn{1}{c}{\textbf{Apple Health SUS:} 63\ \MiniBars{23,17,27,17,27,13,23,10,27,17}} \\
\addlinespace
\midrule

\multirow{7}{*}{Kamala}
& Task 1 & Add a new medication. &
\multicolumn{1}{c}{Partial} & \multicolumn{1}{c}{3.17} &
\multicolumn{1}{c}{Partial} & \multicolumn{1}{c}{2.67} \\
& Task 2 & Record a medication taken. &
\multicolumn{1}{c}{Partial} & \multicolumn{1}{c}{2.83} &
\multicolumn{1}{c}{Partial} & \multicolumn{1}{c}{3.50} \\
& Task 3 & Check today’s medication schedule (daily reminder). &
\multicolumn{1}{c}{Yes} & \multicolumn{1}{c}{5.00} &
\multicolumn{1}{c}{Yes} & \multicolumn{1}{c}{3.50} \\
& Task 4 & Reschedule a medication time in the plan. &
\multicolumn{1}{c}{No} & \multicolumn{1}{c}{2.50} &
\multicolumn{1}{c}{Partial} & \multicolumn{1}{c}{3.00} \\
& Task 5 & Share the medication-taken record. &
\multicolumn{1}{c}{No} & \multicolumn{1}{c}{1.00} &
\multicolumn{1}{c}{Partial} & \multicolumn{1}{c}{3.00} \\
& Task 6 & In Settings, change app language to preferred language (Sinhala/Sri Lanka). &
\multicolumn{1}{c}{No} & \multicolumn{1}{c}{1.00} &
\multicolumn{1}{c}{No} & \multicolumn{1}{c}{1.50} \\
& \multicolumn{2}{l}{\textbf{System Usability Scale, 0--100}} &
\multicolumn{1}{c}{} &
\multicolumn{1}{c}{\textbf{Medisafe SUS:} 34\ \MiniBars{10,30,10,33,27,17,13,23,13,33}} &
\multicolumn{1}{c}{} &
\multicolumn{1}{c}{\textbf{Apple Health SUS:} 44\ \MiniBars{25,15,23,30,28,10,25,15,30,18}} \\
\bottomrule
\end{tabular}%
}
\parbox{0.95\linewidth}{\footnotesize\emph{Note.} \textbf{Completed} is the adjudicated outcome (\emph{Yes/Partial/No}). \textbf{Ease of Use} is the mean 1--5 rating across evaluators (1=hardest, 5=easiest). SUS scores are reported on the standard 0--100 scale.}
\end{table}

\begin{figure*}[!ht]
\centering
\includegraphics[width=\textwidth]{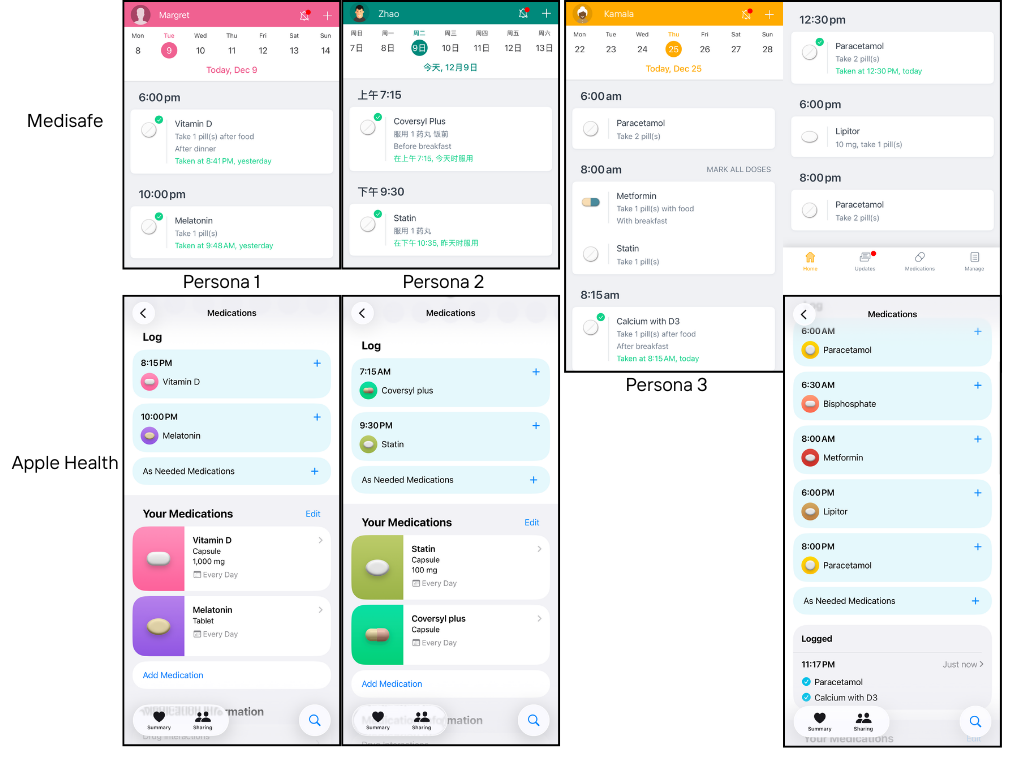}
\caption{Comparison of CW outcomes across three personas (P1, P2, P3) in Medisafe (top row) and Apple Health (bottom row). The figure shows the actual UI states encountered during task execution, highlighting persona-specific differences in navigation ease, visual load, and interaction success.}
\label{fig:persona_ui}
\end{figure*}

Table \ref{tab:cw_tasks_eval_all} summarises the results of Cognitive Walkthrough, which revealed a consistent usability gradient across personas, with \textbf{P1 (Margaret) performing the best, followed by P2 (Zhao), and P3 (Kamala) exhibiting the greatest difficulty} in both MediSafe and Apple Health. Although evaluators' personal experience with each app produced small variations in task-level scores, this gradient appeared across both Medisafe and Apple Health. As illustrated in Figure~\ref{fig:persona_ui}, the differences were visible directly in the UI: Margaret navigated both apps with minimal hesitation, Zhao progressed more slowly—particularly when UI elements required English comprehension or multi-step confirmations—and Persona 3 encountered repeated difficulty locating controls, interpreting system feedback, and completing multi-action tasks such as rescheduling or verifying medication history. This ordering underscores that digital health usability is shaped not simply by age or health status alone, but by their underlying facets. From a health perspective, differences in motivation, tech proficiency, received care, trust and privacy expectations, and health self-efficacy substantially influenced how easily each persona navigated and interpreted system features.

First, the CW results empirically verify that health characteristics significantly shape digital interaction quality, but not in isolation. Across tasks, evaluators attributed Kamala’s moderate performance to a layered interaction of HealthMag facets: lower \textit{motivation} to explore unfamiliar functions, reduced \textit{tech proficiency}, limited \textit{received care}, variable \textit{trust \& privacy} expectations, and weaker \textit{health self-efficacy}. These factors manifested in slower task initiation, greater hesitation when interpreting system feedback, and heightened uncertainty during multi-step tasks such as rescheduling or sharing data. Conversely, Margaret’s strong domain knowledge, higher trust, and greater self-efficacy enabled smooth task completion even when UI complexity increased, illustrating that health facets operate as multiplicative—not additive—determinants of UX.

Second, the CW substantiates the Elderly Age Mag calibration by demonstrating how age-related phenomena concretely disrupt task execution. Kamala’s performance underscored the compounding effects of \textit{visual impairment}, \textit{physical difficulties}, and reduced \textit{tech proficiency}, with evaluators repeatedly noting struggles in locating small interaction targets, navigating multi-panel flows, or manipulating time settings. At the same time, the evaluation supports the calibrated finding that chronological age alone is insufficient as a design predictor: \textit{education} and \textit{received care} emerged as equally influential. Evaluators reported that when adequate support or prior familiarity was present, certain older personas (e.g., Margaret) exhibited fewer age-related constraints than some younger but low-efficacy users described in expert interviews. This confirms the expert panel's earlier assertion that ageing is heterogeneous and interaction barriers arise from configurations of facets rather than from age per se.

Third, the SUS and qualitative CW evidence highlighted a notable difference in application orientation. Medisafe was consistently rated as stronger in functional requirements and medication-specific operations, such as logging doses, handling schedules, and adjusting reminders. Evaluators described it as “more focused,” “more medication-centric,” and “better optimised for task flow,” particularly for personas with stable proficiency. In contrast, Apple Health received higher ratings for non-functional and inclusive design features, including data sharing, ecosystem integration, and language accessibility. Evaluators noted that Apple Health’s lightweight, system-native environment provided a sense of familiarity and reduced cognitive effort for users already embedded in the iOS ecosystem, despite being less specialised in medication workflows. This reinforces the facet-based insight that UX quality emerges from the match between user facets and app design orientation, rather than from an absolute ranking of applications.

Taken together, these findings demonstrate that the CW results not only validate the practical necessity of the HealthMag and Elderly Age Mag models but also reveal how their facet-structured personas manifest in real interaction behaviour. Both facet sets offer explanatory power for why personas diverged in performance and why application suitability varies across user archetypes. The results highlight that successful digital health engagement requires systems that are flexible across multiple facets rather than solutions that rely on narrow assumptions about chronological age or health conditions.

\section{Discussion}\label{sec5}
We have described the development and assessment of an evidence-based Magnifier framework for modelling and evaluating health-related requirements. Our uage of the proposed framework demonstrates its potential value for improving the delivery of better digital-health software used by older adults.

We operationalised interaction-time consequences of health status into a novel framework based on the InclusiveMag concept -- \textit{HealthMag}. This provides a compact facet set with explicit endpoints and walkthrough prompts grounded in a synthesis of 130+ studies. Rather than modelling diagnoses, HealthMag models how health-related differences manifest as actionable facets.

We then calibrated and validated our novel HealthMag through multidisciplinary expert evaluation, using forced-ranking and qualitative feedback to refine facet definitions, remove redundancy, and ensure usability in time-boxed walkthroughs. This process produced a final, discriminative facet set whose endpoints and prompts experts judged interpretable, actionable, and suitable for requirements and evaluation work.

After this, we then paired our HealthMag with an older-adult-calibrated AgeMag lens. This enables more precise attribution of inclusiveness bugs by separating age-driven barriers from health-driven breakdowns and making intersectional failures explicit. We instantiate the dual-lens method with a small set of evidence-grounded personas  that can be used to assess elderly adult-targeted mHealth software requirements and designs.

Finally, we trialled our Elderly HealthMag framework in practice by using it in cognitive walkthroughs of two common medication-management apps. Our findings  illustrate how dual-lens facets explain persona-specific difficulties and guide concrete requirements updates and design remedies to produce better elderly target end user mHealth apps.

\subsection{Actionable Facets: Requirements Modelling with Inclusiveness}
\subsubsection{Design Suggestions for HealthMag facets}
\begin{figure}[!h]
  \centering
  \includegraphics[width=0.8\columnwidth]{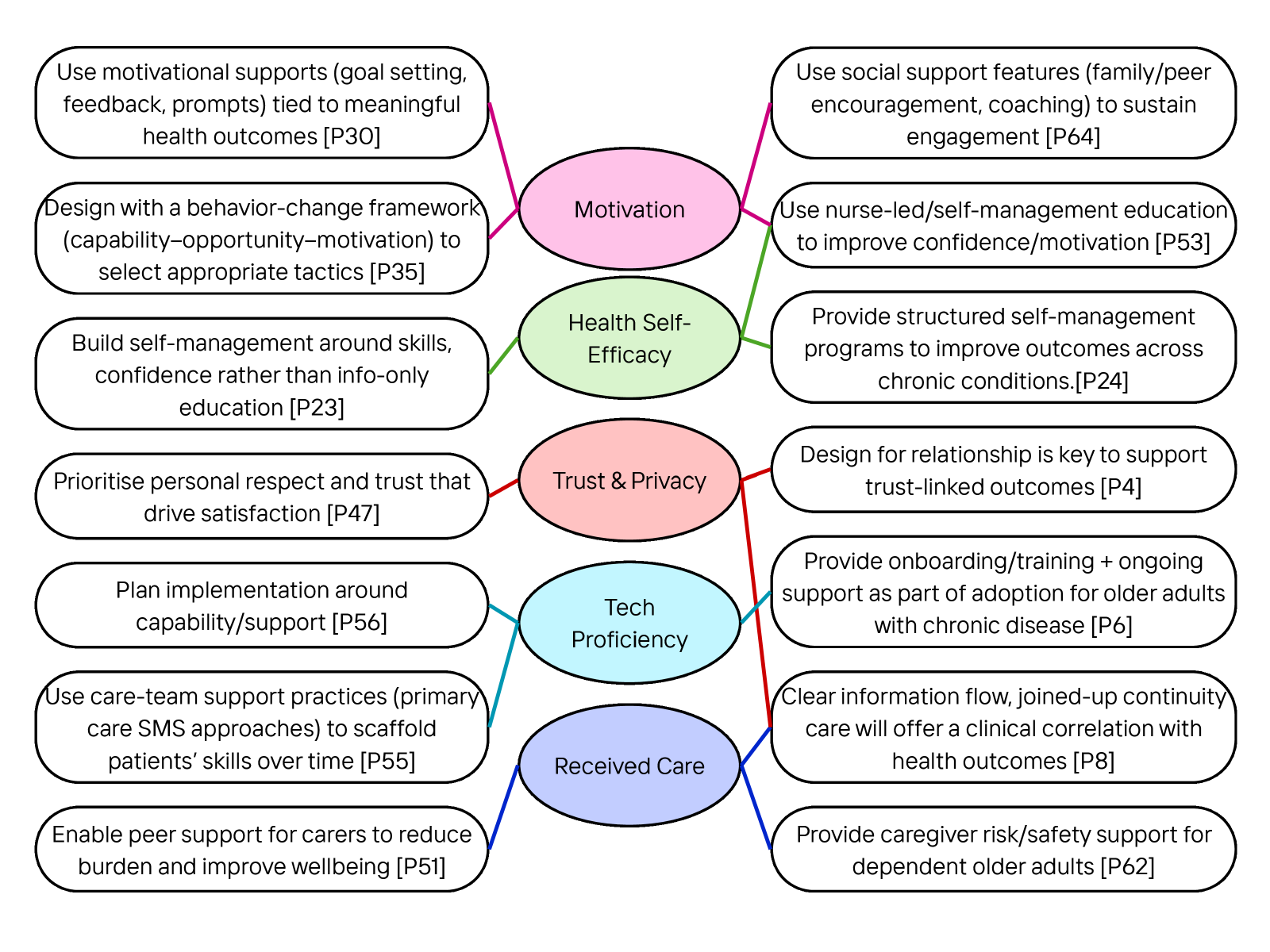}
  \caption{Design recommendations by facet.}
  \label{fig:design-recs-by-facet}
\end{figure}

The literature offers initial guidance on designing health technologies for diverse users. Below we present the design recommendations we identified, organised by five HealthMag facets, and Figure~\ref{fig:design-recs-by-facet} summarises them.
Facet \textit{Motivation} is not purely individual—it is often socially produced and sustained. Zhu et al. emphasise designing for social support (e.g., family/peer encouragement, coaching, shared progress) to maintain health behaviours over time, especially in chronic disease management ~\cite{zhu_relationships_2022}. Michie et al.’s further suggests that “motivation features” should be chosen systematically based on where barriers lie (e.g., capability, opportunity), and paired with appropriate supports such as skill-building or environmental changes~\cite{michie_behaviour_2011}. Finally, Lakerveld et al. highlight concrete tactics—goal-setting, feedback, reminders, and progress visualisation—ideally linked to outcomes users value in daily life (e.g., energy, mobility) rather than biomedical targets alone; this alignment can strengthen engagement and sustain behaviour change~\cite{lakerveld_motivation_2020}.

Facet \textit{Health Self-Efficacy} concerns users’ confidence in managing health tasks and decisions, and can be strengthened through structured, scaffolded support. Huang et al. show that self-management programs combining education, practice, and follow-up improve outcomes, implying designs should go beyond information delivery to skill-building and reinforcement~\cite{huang_effect_2024}. Sun et al. likewise indicate that caregiver-led or self-management education supports health management, motivating tailored guidance ~\cite{sun_impact_2025}. Lorig and Holman further argue that self-efficacy grows through mastery and reinforcement, suggesting achievable goals, feedback, and problem-solving support over time~\cite{lorig_self-management_2003}.

Facet \textit{Tech Proficiency} reflects users’ ability to adopt and use technology, and often depends on training and ongoing support. Bertolazzi et al. note that older adults benefit from onboarding and practical help, suggesting step-by-step guidance and accessible support~\cite{bertolazzi_barriers_2024}. Varsi et al. emphasise matching tools to users’ capability and context, motivating simpler designs that support learning-in-use and routine integration~\cite{varsi_implementation_2019}. Timmermans et al. also highlight care-team scaffolding (e.g., check-ins, troubleshooting) to offset limited proficiency~\cite{timmermans_self-management_2024}.

Facet \textit{Received Care} is influenced by the support users obtain from clinicians, systems, and informal carers, and can be strengthened through coordination and caregiver inclusion. Burch et al. show benefits when care is continuous and coordinated, suggesting better information-sharing and fewer handoff failures~\cite{burch_patient_2024}. Zhang et al. underscore home-care needs for functionally dependent older adults, motivating caregiver risk/safety guidance and decision suppor~\cite{zhang_caregiving_2024}. Peer/befriending support for carers further suggests that supporting carers socially can reduce burden and indirectly improve recipients’ care~\cite{smith_impact_2018}.

Facet \textit{Trust \& Privacy} is shaped by relational continuity, transparency, and coordinated care. Baker et al. associate trust with improved outcomes~\cite {baker_primary_2020}. Burch et al. further suggest trust increases when information follows the patient and services feel joined-up~\cite{burch_patient_2024}. Saultz and Albedaiwi emphasise respect and interpersonal continuity, implying designs should help users feel “known” through respectful communication, consistent follow-up, and clear contact pathways~\cite{saultz_interpersonal_2004}.

\subsubsection{Facet Interdependence}
Consistent with the broader InclusiveMag framework, even when facets are selected carefully to avoid strong correlations, they can still be related in practice. For example, \emph{Education} often correlates with and may even predict \emph{Tech Proficiency}. However, such relationships can themselves become a source of inclusiveness bias if designers assume one facet will compensate for another. In our cognitive walkthroughs, a highly educated, high–tech proficiency user still failed to locate MediSafe’s sharing function due to a lack of familiarity with that specific system, while another user with less education successfully found it.
\subsubsection{AND-Gate Facet Interactions}
Our calibration and walkthrough observations suggest that some facets behave like prerequisites rather than additive contributors, following an ``AND-gate'' pattern. In such cases, usability requires multiple facet conditions to be satisfied simultaneously:
\[
\textit{Usability} = A_{\textit{facet}} \land B_{\textit{facet}}.
\]
For instance, \emph{Motivation} and \emph{Visual Capacity} can act as separate gates: if a user’s eyesight prevents basic reading/targeting, accessibility becomes the blocking failure regardless of downstream support; conversely, if motivation is low, the user may disengage before ever encountering later-stage accessibility barriers or advanced features. This AND-gate logic highlights a common bias in software design: assuming that high-tech proficiency can compensate for poor accessibility (e.g., ``experienced users can figure it out''). In our CWs, users with stronger proficiency still experienced breakdowns when interface signals (labels, confirmations, or language settings) did not match their comprehension needs or when tasks required multi-step commitment under uncertainty. For digital health, this matters because the cost of breakdowns is not only inconvenience but also missed adherence and delayed care.

\subsection{Existing Mag reuse and dual-lens Mag experiments: Extending InclusiveMag’s flexibility}

Adapting a subset of an existing Mag can be both methodologically sound and practically effective. Our results provide some concrete experiences and insights for future Mag development: an existing Mag can be re-scoped to a \emph{smaller} (or \emph{broader}) target population by re-calibrating facet definitions, endpoints, and prompts, rather than rebuilding the method from scratch. In other words, calibration can be used not only to \emph{validate} a facet set, but also to \emph{re-target} it. Concretely, we built \textit{Elderly AgeMag} on InclusiveMag/AgeMag foundations without re-running large-scale literature reviews or new validation surveys~\cite{mendez_gendermag_2019}; instead, we narrowed the scope to older adults in digital-health settings and calibrated the facet set with domain experts so that it better reflects later-life constraints that dominate medication-management tasks (e.g., vision, dexterity, and willingness under uncertainty).

This extends InclusiveMag’s portability claim. InclusiveMag emphasises that Mag mechanics are flexible—facets can be instantiated in flexible personas and applied via specialised walkthrough prompts across different diversity dimensions. Our Elderly AgeMag results suggest an additional, complementary form of flexibility: \emph{facet-set flexibility}. Facets identified in prior Mag work can be reused and re-composed for a new scope (here, shifting from “age across the lifespan” to “older adults in DH”), and then made discriminative again through expert calibration. This goes beyond flexible personas: it increases the \textbf{reusability} of literature-mapped facets in Mag-development studies, because once a facet is evidence-mapped and operationalised, it can serve as a reusable building block that can be re-scoped, re-ranked, and re-validated for new contexts.

Furthermore, as a \textit{dual-lens Mag} experiment, our cognitive walkthrough (CW) results also indicate when this approach is especially suitable: dual-lens Mag methods are a strong fit when target users simultaneously span two diversity dimensions (here, ageing and health status). In such cases, a single lens risks conflating breakdown causes and producing less precise remedies. By pairing HealthMag with an older-adult-calibrated AgeMag, our CW results support clearer attribution (age-driven vs.\ health-driven barriers) and make intersectional failures explicit—breakdowns that often emerge only when both constraints co-occur in the same task flow.

Overall, our experiments suggest a practical pathway for future Mag work: treat evidence-mapped facets as reusable components that can be adapted to different population scopes through calibration, and prefer dual-lens (or multi-lens) compositions when the intended user group is defined by interacting diversity dimensions.

\subsection{Coverage of Personas in InclusiveMag}
Applying InclusiveMag to health raises a practical boundary question: \emph{how far should persona coverage extend along health severity?} Some cases are extremely severe (e.g., ICU patients or users who are unconscious). While it is ethically important to acknowledge these users, including them as direct-interaction personas can impose unrealistic expectations on designers, because software use may be infeasible or entirely proxy-mediated. This tension mirrors a long-standing requirements-engineering challenge: personas must be representative without becoming either overly narrow or implausibly broad.

Our method addresses this boundary problem by delegating coverage decisions to domain experts who routinely work with health-diverse users. Rather than relying on designer assumptions or abstract inclusivity ideals, we asked potential persona users—especially clinicians—to judge which conditions should be meaningfully represented \emph{as direct users} and which should be treated as \emph{caregiver-mediated} or out of scope for interaction-time evaluation. In our results, experts reliably distinguished cases where digital tools can provide reasonable support from those where interaction is indirect or infeasible. Incorporating this domain judgment into persona calibration yields a more balanced representation: it promotes equity while keeping the method actionable, and it makes the coverage boundary explicit and justifiable rather than implicit.  

This approach generalises beyond health: for any diversity dimension, practitioners must make defensible scope decisions about which user groups to represent directly in personas versus treat as context or proxy-mediated use. For example, in culturally and linguistically diverse settings, teams may need to decide which language communities or communication norms to represent explicitly, given the deployment context and available evidence, and which to address through configurable localisation and support pathways. By making these boundary decisions explicit and expert-justified, the method improves both transparency and practical adoptability.

\subsection{Evidence-driven, LLM-generated Personas: A practical way of Adopting LLM in RE}
Recent studies have demonstrated that large language models (LLMs) can efficiently generate personas that appear realistic. For example, PersonaCraft uses LLMs to produce personas by prompting models with demographic and contextual constraints, emphasising narrative plausibility and scalability~\cite{jung2025personacraft}. Similarly, other work shows LLMs can synthesise how user archetypes from textual descriptions or datasets, often focusing on creativity, diversity, or stylistic variation~\cite{lazik2025impostor}. While these approaches show that LLMs are capable persona generators, they largely treat personas as end products of generation, with limited attention to how personas are analytically grounded, calibrated, or systematically aligned with design constructs. As a result, personas risk becoming visually convincing but methodologically opaque artefacts, especially in sensitive domains such as health and ageing.

Our method adopts a fundamentally different stance. Rather than asking the LLM to generate personas freely, we use it as a constrained synthesiser operating over rich, structured inputs. Persona generation is preceded by systematic preparation of requirements, population distribution statistics, and persona structure derived from literature and validated evidence, which are explicitly provided to the model. This helps ensure that the generated personas are grounded in domain-relevant data, not inferred solely from general language patterns. Moreover, personas are constructed using strictly defined and named components—the HealthMag and Elderly AgeMag facets—rather than implicit or emergent attributes. In this way, the LLM does not decide what constitutes a persona; it instantiates personas within an existing conceptual framework. This design choice contrasts with prior LLM persona studies, where persona attributes are often implicit outcomes of prompting strategies rather than explicitly traceable design elements.

A further distinction lies in calibration and iteration. Prior LLM persona approaches typically evaluate personas through face validity or plausibility checks, stopping once the generated artefacts appear reasonable. In our method, the LLM-generated personas (20 candidates) are intentionally treated as intermediate artefacts, which are then reviewed and calibrated by domain experts in health, requirements engineering, and digital health. Feedback explicitly targets coverage boundaries, prevalence, severity, and interaction assumptions (e.g., caregiver-mediated use), leading to revised personas with built-in flexibility (Appendix G). This step is particularly important for elderly health contexts, where technology use is often indirect and strongly shaped by health severity and socio-economic constraints—factors that LLMs alone do not reliably infer. By embedding expert calibration into the pipeline, the method mitigates risks of bias amplification and overgeneralization that have been reported in generative persona work.

Overall, this pipeline works because it integrates LLMs into a theory-driven, evidence-grounded, and expert-calibrated process, rather than positioning them as autonomous persona authors. Compared to existing LLM-generated persona studies, our approach demonstrates how LLMs can be operationalised as reliable support tools for requirements engineering: generating personas that are not only realistic in appearance but also analytically transparent, systematically structured, and suitable for inclusive design reasoning in complex domains.
\subsection{Practical adoption and future work}
\textbf{Adoption guidance.} Elderly HealthMag is intended to be usable as a lightweight workflow: select (or instantiate) facet endpoints via personas, run a structured CW with facet-aware questions, and log breakdowns as traceable inclusiveness bugs linked to facet-driven explanations and concrete remedies. Although we demonstrated the method on medication-management applications, the facets target interaction-time phenomena that recur across digital-health systems (e.g., patient portals, appointment scheduling, monitoring dashboards), making the approach transferable to other DH contexts.

\textbf{Future work.} Three directions follow naturally: (1) extending the Mag family by developing and calibrating new diversity dimensions (e.g., disability severity, cultural identity) using the same evidence-to-calibration pipeline; (2) systematically studying facet-set re-scoping (expanding/shrinking) as a reusable mechanism for adapting existing Mags to new contexts; and (3) evaluating dual- (or multi-) lens compositions at larger scale to characterise when multiple diversity dimensions interact in ways that single-lens methods systematically miss. Together, these directions position Elderly HealthMag as a reusable foundation for inclusive and bias-aware software engineering in digital health.

\section{Threats to Validity}\label{sec6}


\subsection{Development of Mag Facets}
\paragraph{Soundness}
Soundness asks whether the HealthMag and AgeMag facet values vary meaningfully across health conditions and age, and whether such variation plausibly affects software use. Our soundness claim is largely true by construction: candidate facets were derived from a broad, cross‑disciplinary literature base and retained only when supported by evidence linking them to interaction‑relevant outcomes (counterexamples in Appendix ~\ref{sec:appendix_b}). Nevertheless, soundness depends on the quality of the underlying evidence and our synthesis. Potential threats include 1) selection bias introduced by snowballing (e.g., over‑representation of particular conditions or healthcare systems), 2) coding and interpretation bias when mapping constructs to facet endpoints, and 3) publication bias favouring significant results. We mitigated these risks by applying saturation criteria during literature snowballing(forward and backward), multi-disciplinary evidence validation, and conducting in‑depth qualitative calibration with diverse domain experts. Despite these measures, residual error remains possible.
\paragraph{Completeness}
A fundamental tension exists between comprehensive coverage and cognitive manageability. As health status is multifactorial, the literature suggests numerous plausible attributes beyond our selected list. Hence, we intentionally traded exhaustiveness for practicality, optimising for a facet set that is ``small enough to keep in mind, large enough to matter.'' However, this does not guarantee what users of Elderly HealthMag (like Digital health software designers or clinicians) think. Through calibration with domain experts who also represent potential user demographics, we ensured that the final facets capture the primary drivers of user experience.
\paragraph{Construct Validity of Facets and Endpoints.}
Another central threat is whether our facet labels and endpoints mean what practitioners and users would expect them to mean. Health-related constructs can be context-sensitive (condition type, care pathway, culture, language), and an endpoint that appears clear in one setting may be ambiguous in another. Expert calibration improves construct validity but cannot eliminate this threat: experts may disagree, and their judgments may reflect local norms. 

\subsection{Calibration with experts}
Our calibration relies on experts’ judgment. Threats include (i) inconsistent understanding of the facets across experts and (ii) misunderstandings about what a Mag method is and what it is intended to achieve. To mitigate these risks, we assessed experts’ relevant experience prior to the interview and spent time at the beginning of each calibration session establishing a shared understanding of the meaning and intended use of Mag methods. Before asking experts to rank facets, we reintroduced the facets and the HealthMag/AgeMag approach in detail and discussed concrete examples to ensure experts could evaluate the facets from the perspective of intended Mag users.
\subsection{Persona Validity and LLM-Generation Risks.}
Our personas are LLM-generated but evidence-grounded. Still, LLM outputs can introduce unrealistic combinations, missing nuance, or subtle stereotyping. Researcher edits and expert feedback improve plausibility, but personas may remain imperfect proxies for real users. In addition, population-distribution constraints are approximate and may not match the demographics of a specific deployment context.
\subsection{Cognitive Walkthrough}
Our evaluation is based on persona-driven cognitive walkthroughs. As with all analytic inspections, results depend on evaluators’ judgment. Threats include (i) differences in evaluators’ pre-existing experience with the applications, (ii) confirmation bias if evaluators expect certain personas to struggle, and (iii) limited coverage of the walkthroughs. Specifically, we focused on medication-management apps and three dual-Mag elderly personas, rather than conducting broader walkthroughs across additional types of digital-health applications or applying Elderly AgeMag to other categories of software. We mitigated these risks by using structured prompts, maintaining a bug log with on-screen evidence and facet tags, and applying reviewer auditing and consensus reconciliation. Nevertheless, we do not claim causal proof that a given issue is solely due to age or health status; some findings may reflect general usability problems that affect all users. Generalizability is also limited by population scope: Elderly HealthMag targets older adults with chronic conditions and may require recalibration for younger users, acute-care contexts, or non-smartphone environments.

\section{Conclusion}\label{sec7}
This paper shows that health- and age-related inclusivity gaps in digital-health (DH) software can be addressed through a workflow that combines systematic evidence synthesis, expert calibration, and persona-driven cognitive walkthroughs. The result is \textit{Elderly HealthMag}: a dual-lens Magnifier method that models interaction-time requirements and helps detect intersectional bias for older adults living with health conditions.

We first mapped health-related UX determinants for older adults across 134 studies (69 CS/Eng + 65 Health/Psych/SocSci), producing 16 evidence-backed candidate facets. We then operationalised these into a HealthMag model: seven candidate facets refined via expert calibration to five final HealthMag facets. In parallel, we adapted prior AgeMag into an older-adult-specific \textit{Elderly AgeMag} and calibrated it to five final age-related facets. We then consolidated the two lenses into eight non-overlapping \textit{Elderly HealthMag} facets by merging structurally duplicated constructs and preserving complementary health- versus age-specific drivers.

To demonstrate practical use, we instantiated the dual-lens facet endpoints in evidence-grounded personas and applied \textit{Elderly HealthMag} in cognitive walkthroughs of Apple Health (Medications) and Medisafe. These walkthroughs surfaced health-driven barriers, age-driven barriers, and intersectional breakdowns, yielding traceable inclusiveness bugs linked to facet endpoints and actionable redesign priorities. The findings also illustrate that overall usability and facet-specific task support can diverge across applications, reinforcing the need for facet-aware evaluation rather than single-score comparisons. We release reusable artefacts (evidence-to-facet mappings, prompts, and the persona set) to support reuse and replication.

Our conclusions hold under our literature corpus, expert calibration panel, and walkthrough tasks on smartphone medication-management apps using three dual-lens personas; outcomes may differ for other DH domains (e.g., portals, telehealth), younger populations, acute-care contexts, or non-smartphone settings. Next, we will evaluate \textit{Elderly HealthMag} in field deployments and longitudinal studies with product teams, measure comparative effectiveness against baseline methods (time, bug yield, and fix quality), and extend calibration to broader populations and additional DH application categories. We hope \textit{Elderly HealthMag} and its evidence base help the community build DH software that is more inclusive, bias-aware, and requirements-grounded for older adults living with health conditions.

\section*{Acknowledgements}

Xiao and Grundy are supported by ARC Laureate Fellowship FL190100035.

\bibliographystyle{ACM-Reference-Format}
\bibliography{sample-base}










\appendix
\newpage

\section{Papers used for Health Mag Insights}
\label{app_a:papers_for_healthmag}


Table\ref{tab:surveys_for_snowballing_insights} summarises the survey papers that were used for snowballing.

\begin{table}[!ht]
\centering
\renewcommand{\arraystretch}{1.15}
\small
\begin{tabular}{>{\raggedright\arraybackslash}p{8.6cm}
                >{\raggedright\arraybackslash}p{3.2cm}}
\hline
\textbf{Paper} & \textbf{Type} \\
\hline

Human Factors, Human-Centered Design, and Usability of Sensor-Based Digital Health Technologies: Scoping Review (2024)
(\href{https://www.jmir.org/2024/1/e57628}{JMIR})
&
Scoping review
\\

Digital Patient Experience: Umbrella Systematic Review (2022)
(\href{https://www.jmir.org/2022/8/e37952}{JMIR})
&
Umbrella systematic review (review-of-reviews)
\\

Requirements engineering for older adult digital health software: A systematic literature review (2025)
(\href{https://doi.org/10.1016/j.infsof.2025.107718}{ScienceDirect})
&
Systematic literature review
\\

Functional Requirements for Medical Data Integration into Knowledge Management Environments: Requirements Elicitation Approach Based on Systematic Literature Analysis (2023)
(\href{https://www.sciencedirect.com/org/science/article/pii/S1438887123000948}{ScienceDirect})
&
Systematic literature review
\\

Challenges With Developing Secure Mobile Health Applications: Systematic Review (2021)
(\href{https://pmc.ncbi.nlm.nih.gov/articles/PMC8277314/}{PMC})
&
Systematic literature review
\\

The nonfunctional requirement focus in medical device software: a systematic mapping study and taxonomy (2017)
(\href{https://link.springer.com/article/10.1007/s11334-017-0301-6}{Springer})
&
Systematic mapping study
\\

Functional and Nonfunctional Requirements of Virtual Clinic Mobile Applications: A Systematic Review (2024)
(\href{https://onlinelibrary.wiley.com/doi/pdf/10.1155/2024/7800321?utm_source=chatgpt.com}{Wiley PDF})
&
Systematic literature review
\\

\hline
\end{tabular}
\caption{Review-type papers on UX/HCD/usability/requirements engineering in digital health for snowballing.}
\label{tab:surveys_for_snowballing_insights}
\end{table}

\sloppy
\begin{footnotesize}
\end{footnotesize} 
\newpage

\section{16 Candidate Facets Definitions, Examples, and Frequency}
\label{sec:appendix_b}

The following table\ref{tab:facet_definitions} outlines the definitions, illustrative examples, and frequency of facets extracted from the source literature.

\begin{table}[!ht]
    \centering
    \footnotesize 
    \renewcommand{\arraystretch}{1.2} 
    \caption{Definitions, examples, and frequency of extracted facets.}
    \label{tab:facet_definitions}
    \tiny
    \begin{tabular}{@{} p{0.12\textwidth} p{0.24\textwidth} p{0.24\textwidth} >{\centering\arraybackslash}p{0.06\textwidth} >{\RaggedRight\arraybackslash}p{0.24\textwidth} @{}}
        \toprule
        \textbf{Facet} & \textbf{Definition} & \textbf{Examples} & \textbf{Count} & \textbf{References} \\
        \midrule
        
        Received Care & 
        The extent and quality of assistance a patient obtains from formal (providers) and informal (family) networks. & 
        Support from a spouse in scheduling; check-ins from a nurse; access to home aides. & 
        44 & 
        P1, P3, P4, P5, P8, P10, P13, P17, P18, P19, P20, P21, P22, P25, P27, P29, P30, P32, P33, P34, P36, P43, P46, P47, P51, P52, P54, P55, P58, P59, P60, P64, P65, S9, S11, S18, S28, S32, S40, S50, S55, S60, S63, S65 \\
        \midrule

        Technological Proficiency & 
        Ability to seek, find, understand, and appraise health information from electronic sources and apply it. & 
        Competence in using apps; understanding online medical terms; comfort with video calls. & 
        35 & 
        P2, P6, P16, P22, P40, P45, P56, P57, P59, P61, S4, S5, S6, S10, S11, S12, S14, S17, S19, S22, S33, S34, S39, S43, S44, S45, S46, S52, S54, S55, S58, S60, S63, S64, S65 \\
        \midrule

        Trust & 
        Belief that the digital platform and entities will act in the patient’s best interest regarding security and validity. & 
        Confidence in data security; belief in accuracy of symptom checkers; faith in ethical standards. & 
        33 & 
        P2, P3, P4, P5, P6, P11, P12, P17, P18, P19, P20, P23, P40, P43, P45, P47, P50, P51, P52, P54, P56, P62, S1, S19, S21, S30, S31, S35, S37, S39, S45, S50, S56 \\
        \midrule

        Motivation & 
        Internal/external drivers that initiate and maintain health management behaviors. & 
        Desire to avoid complications; incentives for exercise; setting personal health goals. & 
        27 & 
        P1, P6, P7, P15, P20, P24, P26, P35, P36, P48, P49, P50, P53, P54, P57, P62, S8, S12, S18, S24, S25, S26, S34, S38, S43, S54, S63 \\
        \midrule

        Health Self-Efficacy & 
        Belief in one's capacity to execute behaviors necessary to manage their own health. & 
        Confidence in managing glucose levels; belief in ability to quit smoking; capability to use devices. & 
        26 & 
        P1, P3, P6, P13, P21, P24, P26, P28, P29, P30, P32, P35, P36, P39, P53, P54, P55, P56, P62, P63, P64, S2, S20, S36, S38, S64 \\
        \midrule

        Comorbidities & 
        Co-occurrence of two or more chronic conditions, complicating intervention strategies. & 
        Diabetes with Depression; Hypertension with Kidney Disease. & 
        13 & 
        P1, P3, P4, P10, P25, P27, P36, P37, P39, P52, P60, S17, S22 \\
        \midrule

        Living Situation & 
        Physical and social characteristics of a home setting influencing engagement. & 
        Lack of home internet; multi-generational household; shared devices. & 
        13 & 
        P1, P6, P36, P41, P61, P62, S14, S20, S30, S32, S48, S53, S57 \\
        \midrule

        Loneliness & 
        Subjective distress resulting from a discrepancy between desired and actual social relationships. & 
        Feeling isolated despite contacts; low frequency of interactions. & 
        11 & 
        P7, P65, S1, S4, S14, S19, S23, S42, S62, S64, S69 \\
        \midrule

        Societal Acceptance & 
        Degree to which a condition or tech usage is normalized or stigmatized. & 
        Stigma of mental health apps; resistance to vaccination; cultural norms. & 
        10 & 
        P29, P61, P63, S14, S23, S24, S37, S53, S61, S62 \\
        \midrule

        Gender Identity & 
        Personal sense of gender intersecting with access, bias, and usage patterns. & 
        Symptom differences by gender; tailored advice; access barriers. & 
        9 & 
        P1, P4, P15, P16, P37, P41, P44, S14, S49 \\
        \midrule

        Socioeconomic Status & 
        Measure of work experience, economic access to resources, and social position. & 
        Income level; educational attainment; employment status. & 
        9 & 
        P1, P4, P6, P15, P16, P35, P40, P41, P49 \\
        \midrule

        Peer-support Capabilities & 
        Functionality to facilitate reciprocal support among people with similar conditions. & 
        Online forums; mentorship; group chats. & 
        8 & 
        P6, P11, P24, P31, P32, P33, P38, P51 \\
        \midrule

        Empathy & 
        Extent to which a user feels understood and validated by the intervention. & 
        Non-judgmental language; active listening in video consults. & 
        7 & 
        P6, P12, P30, S15, S41, S58, S61 \\
        \midrule

        Cognitive Function & 
        Level of mental process capability dictating usability requirements. & 
        Memory issues; need for simplified instructions; attention span. & 
        7 & 
        P1, P6, P7, P52, S17, S40, S58 \\
        \midrule

        Cultural Competency & 
        Linguistic, cultural, and ethnic factors necessitating tailored interventions. & 
        Native language availability; dietary customs; traditional practices. & 
        4 & 
        P31, P38, P41, P61 \\
        \midrule

        Continuity of Care & 
        Degree to which healthcare events are experienced as coherent and connected. & 
        Seamless data transfer; consistent advice; clear follow-up. & 
        1 & 
        P58 \\
        
        \bottomrule
    \end{tabular}
\end{table}
\newpage

\section{7 Chosen Facets of HealthMag}
\label{sec:appendix_c}
\textbf{Received Care} The "Received Care" facet captures the extent and quality of assistance a patient obtains from formal sources (e.g., healthcare providers), informal networks (e.g., family), and the broader ecosystem (e.g., government services). We prioritized this facet for two reasons: (1) the level of received care significantly moderates an individual's accessibility to and capability with digital health software; and (2) specific health conditions influence the volume of care a person can elicit from the ecosystem~\cite{albrecht_adherence_2016,baker_primary_2020,barker_association_2017,gray_continuity_2018,haggerty_experienced_2013,hansen_continuity_2013,hemmings_improving_2019, kripalani_reducing_2014,lakerveld_motivation_2020,lorig_self-management_2003,mallick_multivariable_2021}.

This facet encompasses three primary dimensions. The first involves the availability of social care, including professional care networks and social workers. Zhu et al. note that social support directly improves "patient activation" (PAM scores), suggesting that integrating family and friends into the app ecosystem boosts the patient's own capability~\cite{zhu_relationships_2022}. However, Gray et al. caution that caregivers and professionals may underestimate patient abilities as conditions worsen, leading to over-assistance that erodes autonomy~\cite{gray_continuity_2018}. The second dimension is family-oriented support, such as assistance from a spouse, children, or siblings who facilitate device setup, reminders, and escalation. Sorgalla et al. identified that for individuals who are not physically active, support from the broader family network is a key requirement~\cite{sorgalla_improving_2017}. The third dimension involves peer support, such as friends or neighbours who are geographically close and available for assistance. While peer input can enhance coping mechanisms and confidence, it also carries the risk of introducing misinformation~\cite{gray_continuity_2018}. Our evidence highlights significant variance in this area: while some older adults benefit from proactive support (e.g., adult children configuring devices), others experience isolation and fragmented care. Design assumptions that “someone else will help” (e.g., with printing instructions or reconciling medications) unfairly disadvantage those with limited networks. Conversely, systems that over-rely on proxies can undermine users capable of independent action. These findings imply that: (1) systems should explicitly support both self‑management and proxy interactions through configurable roles and permissions; and (2) critical interactions should never strictly require an invisible helper, but should instead offer optional, user‑controlled mechanisms to invite support.

\textbf{Motivation} The Motivation facet captures a user’s willingness to initiate and persist with health‑related software over time, whether for monitoring vital signs, post-surgical rehabilitation, or medication management. We selected this facet because (1) an individual's health condition fundamentally impacts their motivation to use digital health tools, and (2) motivation acts as a primary determinant in the decision to adopt specific software.
First, motivation is heavily influenced by the feedback mechanisms inherent in the software, which help maintain user connection~\cite{albrecht_adherence_2016,bertolazzi_barriers_2024,borji_investigating_2018,eisner_influence_2010,haggerty_experienced_2013,kao_association_2019}. For instance, Rodriguez et al. found that in tele-rehabilitation, immediate feedback serves as the core driver of motivation; patients require confirmation that they are executing exercises correctly to sustain engagement~\cite{cleland_contextualizing_2015}. Second, physical health status directly mediates the will to stay healthy. Jackson et al. stated that pain intensity negatively affects the relationship between motivation and function, with high pain levels lowering motivation~\cite{jackson_arthritis_2020}. This aligns with our previous survey study, where older adults reported feeling “too tired,” “discouraged,” or unable to “see the point” due to conditions such as post-surgical pain~\cite{jmirsurveyxiao}. These observations suggest that motivation is not a stable trait but is dynamically shaped by how software frames tasks, reflects progress, and responds to lapses. Consequently, two design imperatives emerge: (1) workflows that visualise progress early and normalise “slips” with easy re‑engagement paths are essential for supporting users with fragile motivation; and (2) features that anchor tasks to personally meaningful outcomes (e.g., maintaining independence) are more effective at sustaining engagement than purely clinical metrics.
\textbf{Health Self‑efficacy} The Health Self-Efficacy facet reflects an individual's belief in their capacity to manage health tasks, navigate care processes, and execute the behaviours necessary to achieve specific health outcomes~\cite{cleland_contextualizing_2015,albrecht_adherence_2016,hansen_continuity_2013,timmermans_self-management_2024,varsi_implementation_2019,panagioti_self-management_2014,bertolazzi_barriers_2024,bellandi_design_2021,lakerveld_motivation_2020}. Meta-analyses indicate that higher health self-efficacy predicts improved self-management and better outcomes across chronic conditions; conversely, low self-efficacy is associated with avoidance, anxiety, and disengagement, even when the individual possesses adequate theoretical knowledge. While related concepts such as "health literacy" and "health proficiency" describe technical capability, we consolidated these under the term "health self-efficacy." This decision stems from the observation that behaviour change in technology usage is driven less by the static capability to process information and more by the user's \textit{belief} in that capability, which varies significantly based on education and life experience. Empirical evidence supports this distinction; for example, self-management interventions have been shown to be more effective than routine care in managing chronic diseases, significantly improving patients’ quality of life and self-efficacy while reducing depressive symptoms~\cite{huang_effect_2024}. Similarly, Jackson et al. report that in cases of OA/RA, higher self-efficacy correlates with better pain and functional outcomes~\cite{jackson_arthritis_2020}. In our systematic literature review (SLR), primary studies revealed that older adults with low efficacy often expressed fear of "getting it wrong" when adjusting dosages or sending messages, interpreting errors or unclear feedback as evidence that they “should not be doing this alone”~\cite{xiao2025requirements,ferreira_elderly_2014,teixeira_design_2017}. These patterns suggest two critical design insights: (1) interfaces should provide scaffolded, low‑risk “practice” environments with clear, non‑blaming error recovery mechanisms; and (2) timely, specific feedback that acknowledges successful steps—rather than focusing solely on failure—can incrementally build self‑efficacy and reduce abandonment.
\textbf{Cognition}
The Cognition facet addresses the dynamic load on a user’s attention, working memory, comprehension, and decision-making capabilities~\cite{albrecht_adherence_2016,bertolazzi_barriers_2024,borji_investigating_2018,starfield_contribution_2005,van_hoof_what_2016,lindsay_empathy_2012}. Unlike baseline intelligence, this facet reflects the fluctuating cognitive bandwidth often compromised by chronic conditions, medication side-effects, pain, or acute anxiety. Participants in the primary studies frequently described episodes of "losing track" during multi-step tasks, forgetting previous inputs, or feeling overwhelmed by dense medical terminology and simultaneous decision points. These cognitive fluctuations directly impact safety and usability, affecting whether older adults can successfully navigate clinical workflows, retain critical instructions, or recognise when a task is complete.

The literature supports two key design implications: (1) interfaces must employ chunked, linearized workflows with explicit state indicators ("you are here" cues) and minimal requirements for backtracking to support users under high cognitive load; and (2) systems should provide external memory aids—such as persistent summaries, checklists, and interaction histories—to mitigate forgetfulness and eliminate the burden of re-entering or re-inferring information.

\textbf{Comorbidities}
The Comorbidities facet captures how multiple co-occurring conditions compound sensory/motor constraints—specifically vision, hearing, dexterity, mobility, and stamina—thereby dictating interaction possibilities\cite{albrecht_adherence_2016,baker_primary_2020,,barker_association_2017,cho_association_2015,gray_continuity_2018,kao_association_2019}. For example, older adults frequently reported that standard interface elements (e.g., small touch targets, low contrast, fixed font sizes) or prolonged interaction sessions became prohibitive when exacerbated by pain, fatigue, or the need to manage assistive devices like walkers or hearing aids ~\cite{hussey_continuity_2014,njoku_risk_2020,ozorio_autoimmune_2007,wensing_continuity_2021}. These comorbidities extend beyond compliance with basic accessibility standards; they determine whether users can reliably complete time-critical health tasks, such as acknowledging medication reminders or reporting acute symptoms.

Consequently, the evidence points to two essential insights: (1) multimodality—including scalable text, high-contrast modes, and redundant voice/keyboard alternatives—must be treated as a core architectural requirement rather than an optional setting; and (2) interaction patterns should minimize physical effort and explicitly support "pause and resume" functionality to accommodate users whose physical stamina fluctuates throughout the day.

\textbf{Trust}
The Trust facet captures the user’s willingness to rely on the system and its associated actors, heavily influenced by concerns regarding privacy, data security, and the perceived legitimacy of automated recommendations~\cite{cunningham_health_2012,dambhamiller_association_2019,zhang_caregiving_2024,saultz_interpersonal_2004,bazemore_higher_2018}. For older adults, trust is not a static attribute but a dynamic state shaped by prior experiences with healthcare institutions, exposure to misinformation, and interface signals such as branding and endorsements. Our analysis highlights instances where users refused to input sensitive health data or adhere to digital interventions because they "didn't know who was behind the app" or feared data misuse, despite valuing the tool's potential utility. Conversely, visible associations with known clinicians and transparent explanations for algorithmic advice were found to foster the willingness to act.

From these findings, we derive two insights: (1) transparent provenance (disclosing the who, why, and when of information) and plain-language explanations of data usage are absolute prerequisites for informed consent and sustained engagement; and (2) design strategies must treat trust as a renewable resource that is earned through consistent, verifiable cues, rather than assuming a baseline level of user confidence.

\textbf{Tech Proficiency}
The Tech Proficiency facet describes a user's experience with and comfort using digital devices, encompassing basic operational skills, mental models of system behaviour, and confidence in error recovery~\cite{umm_e_mariya_shah_usability_2022,hendriks_designing_2013,maria_luisa_rodriguez-almendros_design_2021,bertolazzi_barriers_2024,jussli_senior_2021,robinson_participatory_2022}. The spectrum of literacy in our corpus was broad: while some older adults possessed extensive history with smartphones, others were late adopters who restricted usage to narrow functions like calling. Low proficiency often manifested as navigation uncertainty, a fear of “breaking” the device, and difficulty interpreting system feedback, whereas high-proficiency users demonstrated a willingness to explore and customise tools. Crucially, this facet intersects with health status; fatigue, pain, or cognitive fluctuations can temporarily degrade effective literacy even in experienced users.

These observations suggest two design priorities: (1) onboarding processes and critical workflows must accommodate diverse experience levels through clear signposting, optional “how it works” overlays, and forgiving, reversible actions; and (2) error-handling mechanisms should reassure users that mistakes are safe and repairable, rather than assigning blame, thereby encouraging learning and exploration over time.
\newpage

\section{Themes Derived from SLR and Data-driven Personas with LLM}
\label{sec:appendix_d}

\subsection{Themes Derived from SLR}
The following themes were identified through the systematic literature review (SLR) and served as the basis for persona generation.

\begin{description}
    \item[Theme 1: Medication Support] \hfill \\
    \textbf{Description:} Providing comprehensive support for medication-related needs, including reminders, multi-drug interaction alerts, and management of comorbidities. It ensures safer, more effective medication use and adherence to prescribed treatment plans. \\
    \textbf{Origin Words:} ``Correct and timely drug intake assistance to avoid multi-drug interactions''; ``The application should provide medication alerts to remind users about medication schedules.''; ``Reminders for the exercises, the medicine and the meetings are sent to Mr. Pieters through Julie...''; ``Vitamin D, calcium, or combined supplementation...'' \\
    \textbf{Personas (2):} 1, 3, 4, 17.

    \item[Theme 2: Health and Welfare Monitoring] \hfill \\
    \textbf{Description:} Elderly life monitoring and support, including home living, health, daily activities, social interactions, disease management, medical records, healthcare communication, and advice. \\
    \textbf{Origin Words:} ``A COPD nurse uses the two-way video to check-up...''; ``The quality of life among the elderly can be affected by place of residence...''; ``Chronic obstructive pulmonary disease: screening'' \\
    \textbf{Personas (6):} 2, 6, 7, 18, 19, 20.

    \item[Theme 3: Social Interaction and Emotional Care] \hfill \\
    \textbf{Description:} This theme involves communication and interactions between older adults and their family or friends, participation in social activities, emotional states (e.g., loneliness, happiness), and challenges or emotions faced by caregivers. \\
    \textbf{Origin Words:} ``Communication way between parent and child, Maintaining social contacts in the neighbourhood''; ``Julie suggests activities and new inhabitants for him to meet...''; ``All of these programs allow for the family to see mom or dad...'' \\
    \textbf{Personas (4):} 9, 10, 11, 12.

    \item[Theme 4: Emergency Support and Fall Prevention] \hfill \\
    \textbf{Description:} This theme focuses on providing timely emergency support to older adults, ensuring their safety and well-being in critical situations, contact their caregivers and family when necessary. \\
    \textbf{Origin Words:} ``If an individual cannot make all decisions and needs support...''; ``The holidays give family members of all generations an opportunity to see how one another are doing...''; ``The daily caregiver sent a quick text to the daughter every morning...'' \\
    \textbf{Personas (3):} 5, 8, 13, 14, 16.
\end{description}

\subsection{Detailed Persona Profiles}
Below are the 20 calibrated personas generated based on the themes above.
\includepdf[pages=-, scale=0.65, pagecommand={}]{appendices/20_personas.pdf}
\newpage

\section{Prompts for generating personas with LLM}
\label{sec:appendix_e}
This section outlines our LLM workflow, including the model(s) used, the full prompts, and the iterations we ran to generate personas based on predefined structures, requirements, and evidence data.

\begin{figure} []
    \centering
    \includegraphics[width=0.6\textwidth]{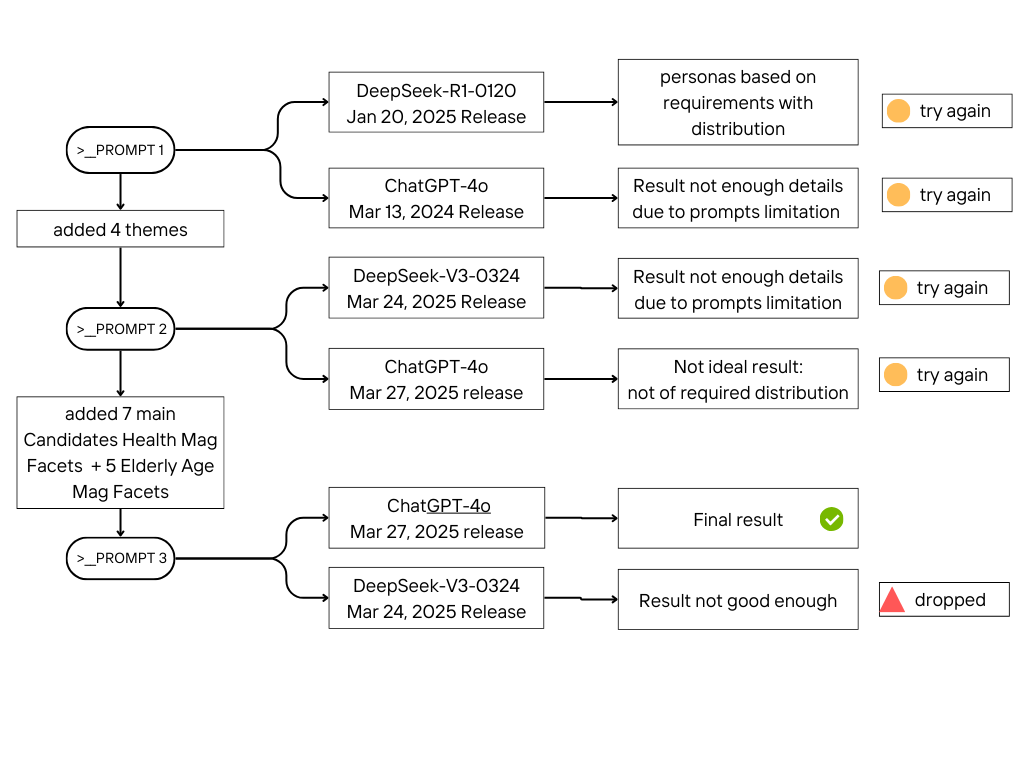}
    \caption{LLM Workflow}
    \label{fig:llm_process}
\end{figure}

\begin{figure} []
    \centering
    \includegraphics[width=0.6\textwidth]{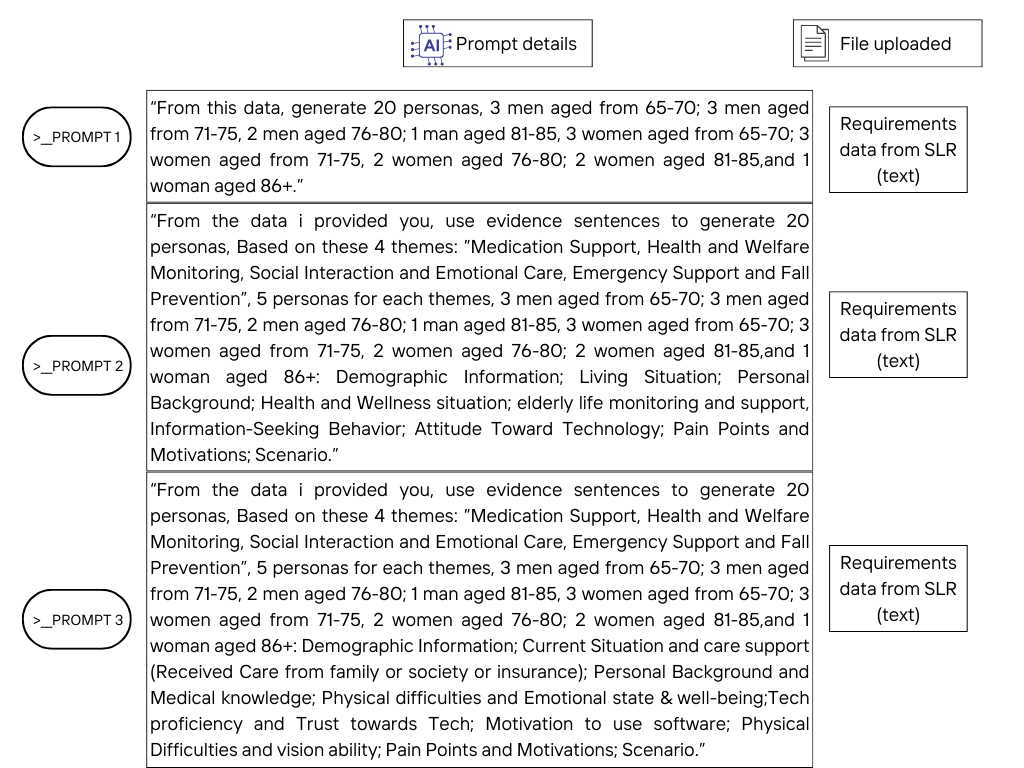}
    \caption{LLM full Prompts}
    \label{fig:llm_prompts}
\end{figure}
\newpage

\section{ElderlyAgeMag}
\label{sec:appendix_f}

Table\ref{tab:barriers_taxonomy_concise} summarises the barriers mapped to Health Mag and Elderly Age Mag Facets.
\begin{table*}[ht]
\centering
\caption{Barriers mapped to Health Mag and Elderly Age Mag Facets}
\label{tab:barriers_taxonomy_concise}
\footnotesize
\renewcommand{\arraystretch}{1.5}
\begin{tabular}{p{2cm} p{6.5cm} p{3cm} p{3cm}}
\toprule
\textbf{Barrier Layer} & \textbf{Evidence (Participant \& Quote)} & \textbf{Health Mag Facets} & \textbf{Elderly Age Mag Facets} \\
\midrule

\textbf{1. Intrinsic \newline (Clinical)} & 
\textbf{Pain:} ``The main issue is the pain, which will certainly affect his ability to use a mobile phone'' (P7). \newline
\textbf{Medication Side-effects:} ``She will see the world all in green'' (P1); ``When he takes the medication, he cannot concentrate'' (P1). \newline
\textbf{Comorbidities:} ``Reduced hand dexterity... they can’t click on all the buttons'' (P2). & 
\textbullet\ Motivation \newline
\textbullet\ Health Self-efficacy & 
\textbullet\ Physical Difficulties \newline
\textbullet\ Visual Impairment \\
\hline

\textbf{2. Interaction } & 
\textbf{Habit/Format:} ``My advisor wanted me to hand out its assignments on paper'' (P3). \newline
\textbf{Memory:} ``Struggle to like, remember where prescriptions are stored'' (P2). \newline
\textbf{Reluctance:} ``Reluctant to trust the features recommended by the platform'' (P9). & 
\textbullet\ Trust  \newline
\textbullet\ Tech Proficiency \newline
\textbullet\ Health Self-efficacy & 

\textbullet\ Tech Proficiency \\
\hline

\textbf{3. Extrinsic \newline (Contextual)} & 
\textbf{Location:} ``Digital health systems often end up helping those who are really well-connected'' (P4). \newline
\textbf{Cost (SES):} ``Extent of software usage depends on whether your software requires a fee'' (P8). \newline
\textbf{Hardware:} Signal quality varies by device (P9). & 
\textbullet\ Received Care \newline
\textbullet\ Motivation & 
\textbullet\ Received Care\newline
\textbullet\ Education
\\
\bottomrule
\end{tabular}
\end{table*}
\newpage

\section{3 personas of ElderlyHealthMag}
\label{sec:appendix_g}

\begin{figure} [h!]
    \centering
    \includegraphics[width=0.8\textwidth]{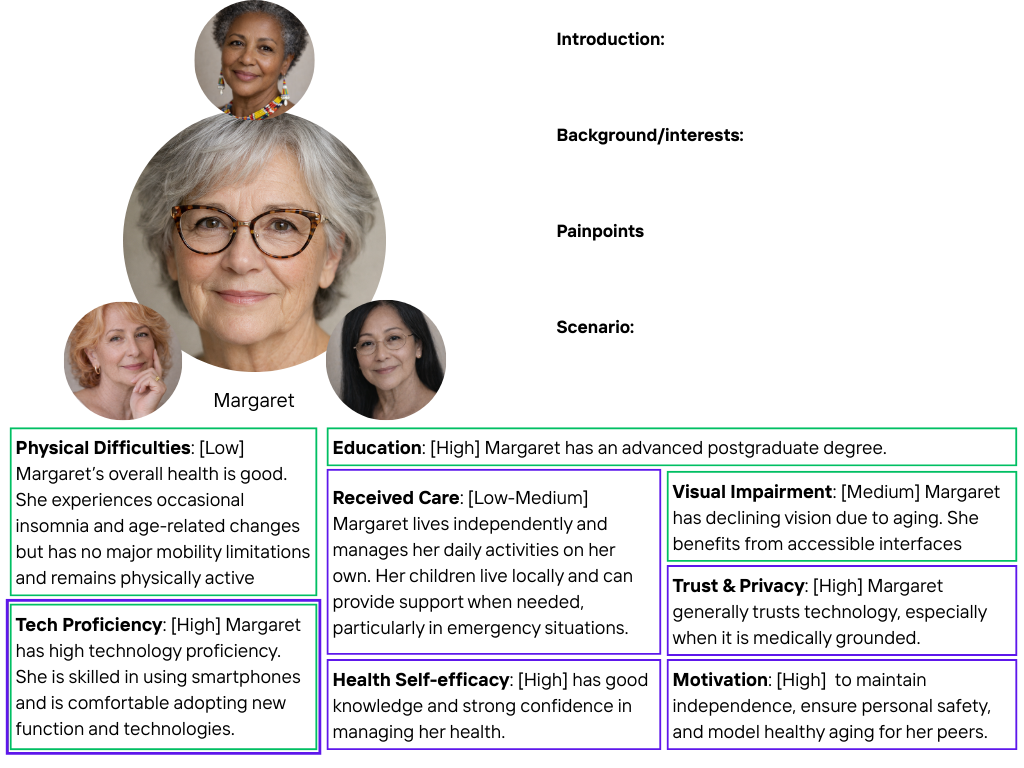}
    \caption{Persona 1: Margaret}
    \label{fig:persona1}
\end{figure}

\newpage

\begin{figure} [h!]
    \centering
    \includegraphics[width=0.8\textwidth]{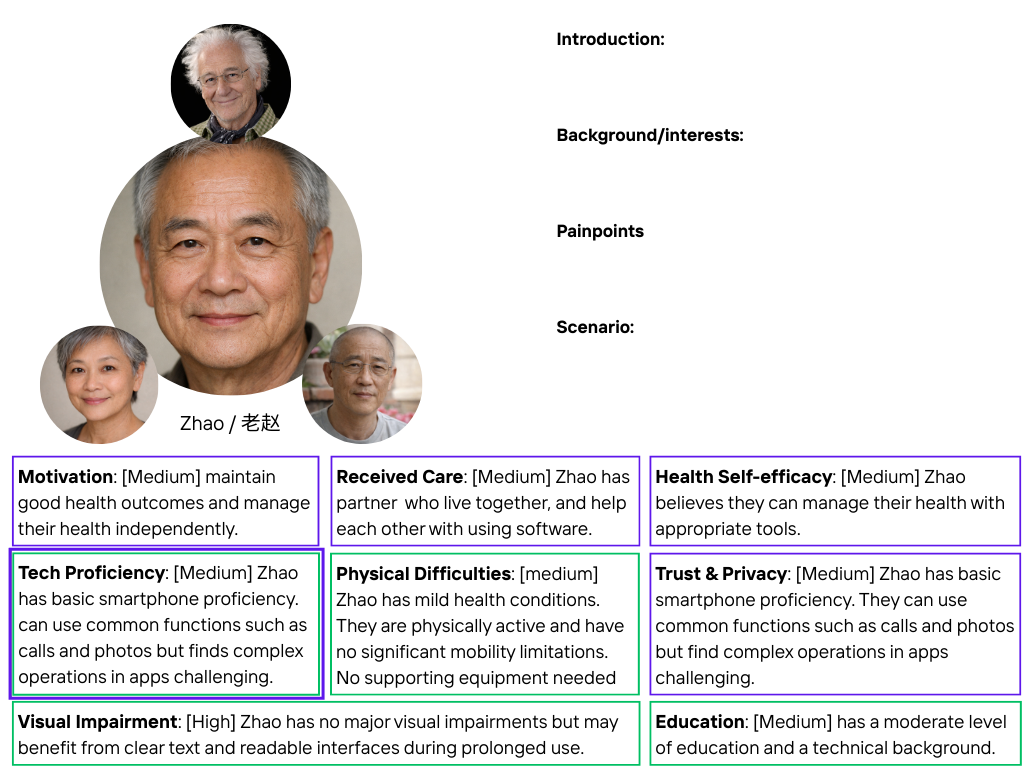}
    \caption{Persona 2: Zhao}
    \label{fig:persona2}
\end{figure}

\newpage

\begin{figure} [h!]
    \centering
    \includegraphics[width=0.8\textwidth]{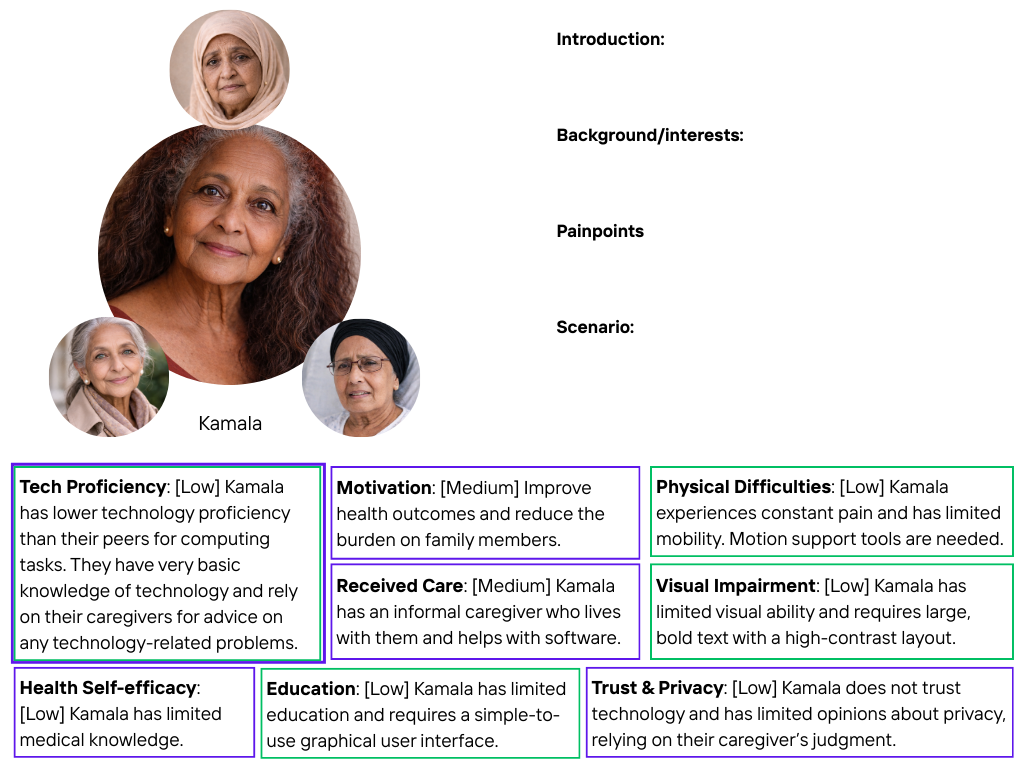}
    \caption{Persona 3: Kamala}
    \label{fig:persona3}
\end{figure}
\newpage

\section{Interview Guideline}
\label{app:interview_guide}
This appendix presents the complete semi-structured interview guideline used during the expert calibration phase, detailing the sequential protocol for validating the necessity, acceptance, and usability of the Health Mag and Age Mag facets and personas.
\includepdf[pages=-, scale=0.55, pagecommand={}]{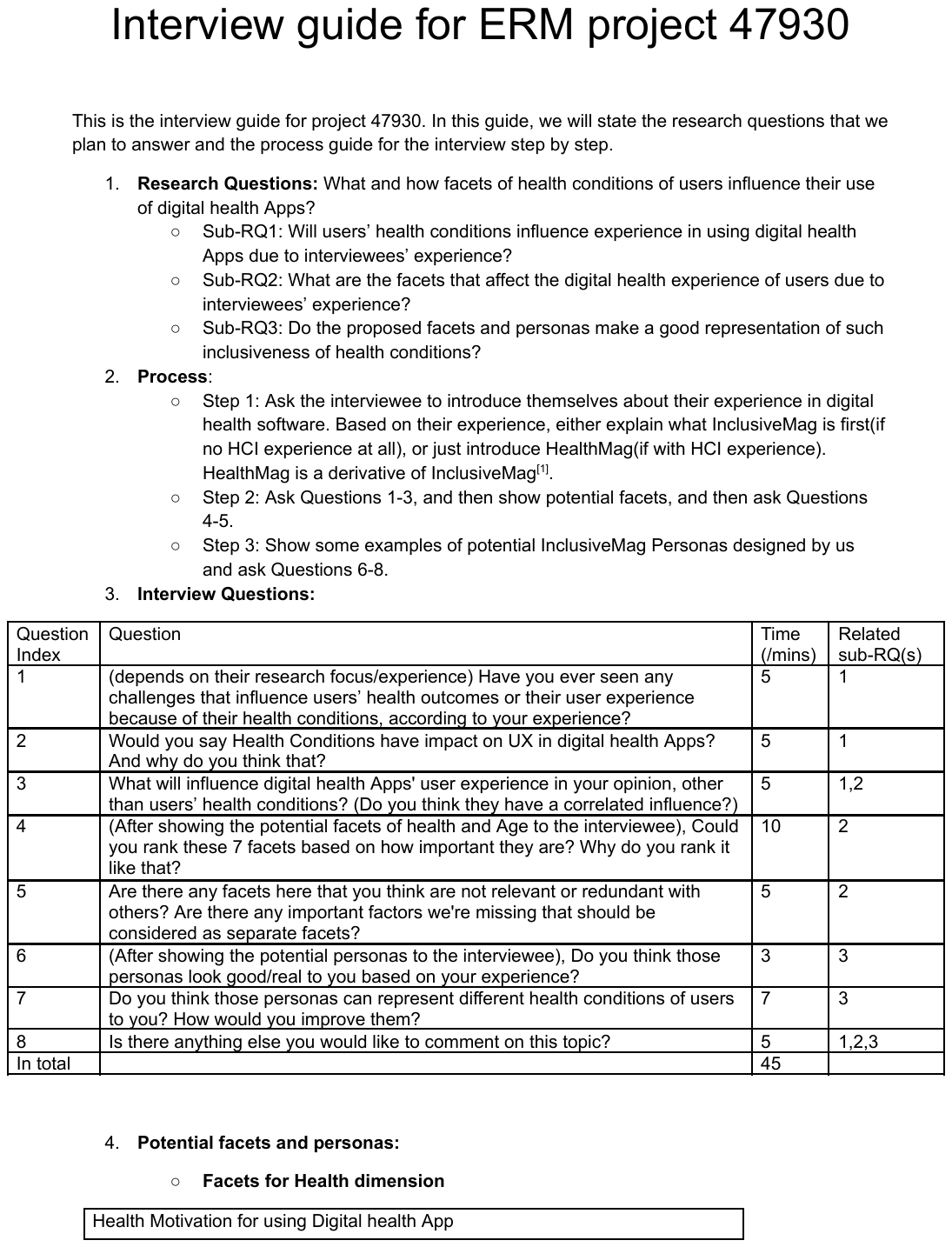}
\newpage

\section{Cognitive Walkthrough}
\label{app:i}

This appendix presents the complete Cognitive Walkthrough details.
\begin{table}[!ht]
\centering
\footnotesize
\caption{Developer Roles, Applications, and Assigned Persona--Task Combinations for Cognitive Walkthrough}
\label{tab:CW_evaluators}
\begin{tabular}{p{1.5cm} p{5.5cm} p{2.3cm} p{6cm}}
\hline
\textbf{Developer} & \textbf{Role / Background} & \textbf{Application} & \textbf{Assigned Personas and Tasks} \\
\hline

\textbf{Evaluator 1} & 
Software developer of medication management apps &
Medisafe \newline Apple Health &
\begin{itemize}
    \item Persona 1: Tasks 3, 5
    \item Persona 2: Tasks 1, 2, 3, 5
    \item Persona 3: Tasks 1, 2, 3, 5
\end{itemize} \\

\textbf{Evaluator 2} & 
Software developer of medication management apps &
Medisafe \newline Apple Health &
\begin{itemize}
    \item Persona 1: Tasks 1, 2, 4, 5
    \item Persona 2: Tasks 4, 5
    \item Persona 3: Tasks 1, 2, 4, 5
\end{itemize} \\

\textbf{Evaluator 3} & 
Software developer of medication management apps &
Medisafe \newline Apple Health &
\begin{itemize}
    \item Persona 1: Tasks 1, 2, 3, 4
    \item Persona 2: Tasks 3, 4
    \item Persona 3: Tasks 1, 2, 3, 4
\end{itemize} \\

\textbf{Evaluator 4 \newline(Author 1)} & 
PhD Candidate in Software Engineering; developer of digital health apps &
Medisafe \newline Apple Health &
\begin{itemize}
    \item Persona 1: Tasks 1, 2
    \item Persona 2: Tasks 1, 2, 6
    \item Persona 3: Tasks 1, 2, 6
\end{itemize} \\

\hline
\end{tabular}
\end{table}

\begin{figure} [!t]
    \centering
    \includegraphics[width=0.7\textwidth]{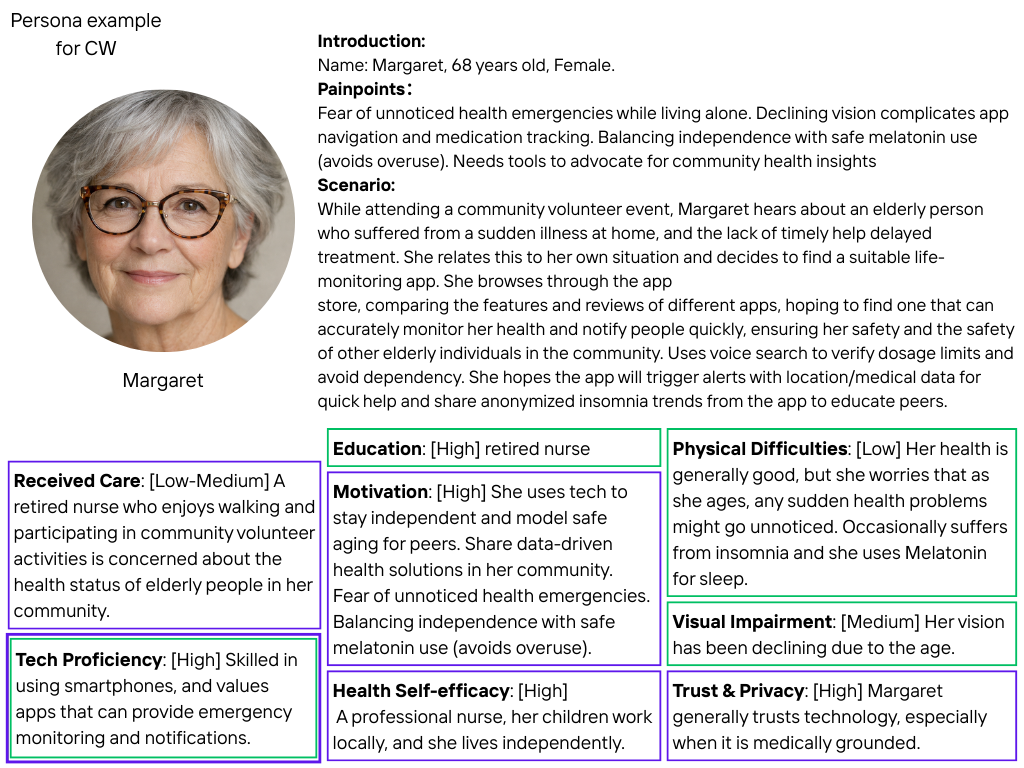}
    \caption{Persona 1: Margaret}
    \label{fig:persona1_forCW}
\end{figure}

\begin{figure} [!t]
    \centering
    \includegraphics[width=0.7\textwidth]{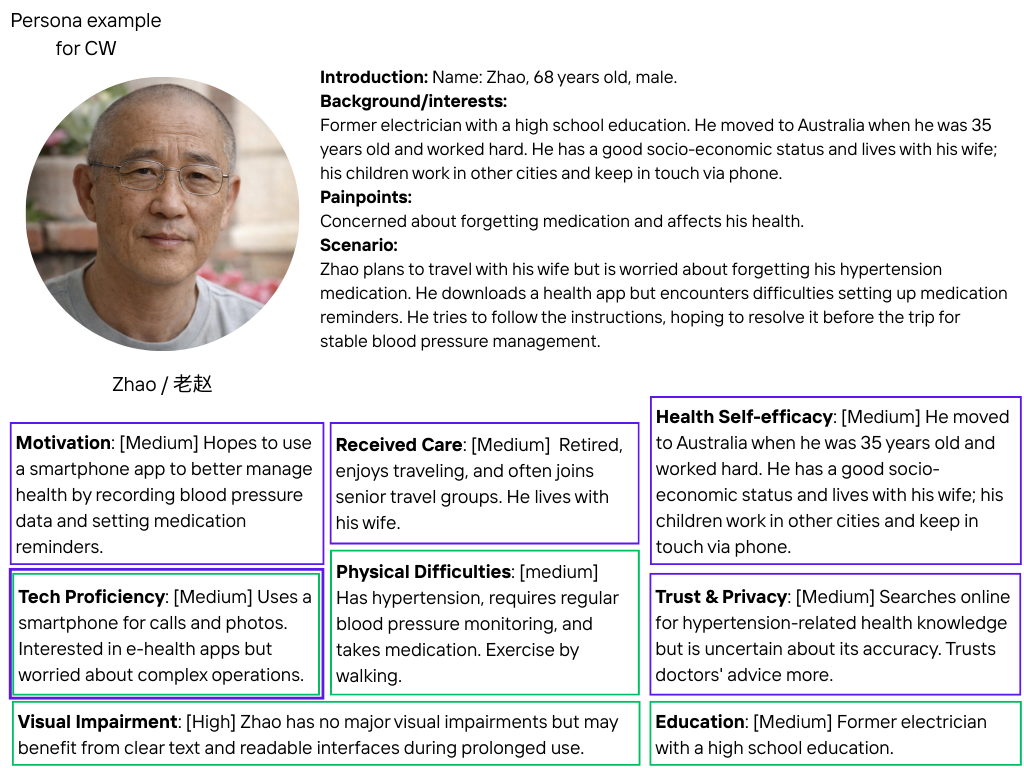}
    \caption{Persona 2: Zhao}
    \label{fig:persona2_forCW}
\end{figure}

\begin{figure} [!t]
    \centering
    \includegraphics[width=0.7\textwidth]{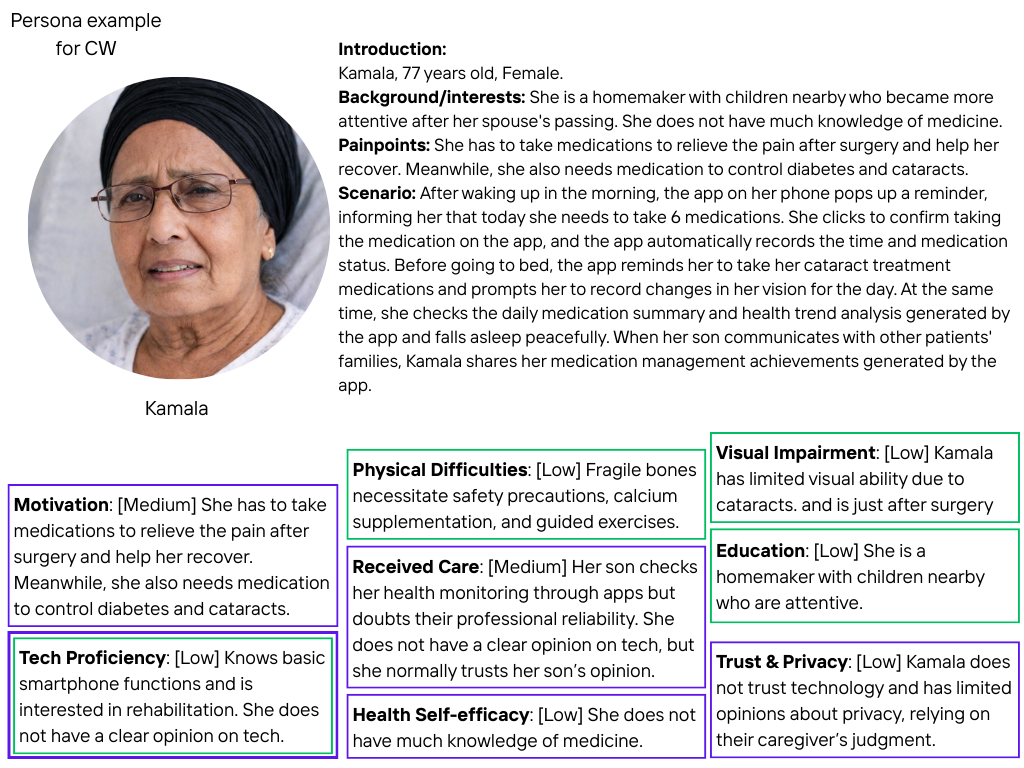}
    \caption{Persona 3: Kamala}
    \label{fig:persona3_forCW}
\end{figure}

The personas we used for the Cognitive Walkthrough are based on our proposed Elderly HealthMag, specifically implemented for the target applications, as shown in Figure~\ref{fig:persona1_forCW}, \ref{fig:persona2_forCW}, \ref{fig:persona3_forCW}. 

\end{document}